\documentclass[11pt,a4paper,american]{extarticle}
\usepackage[T1]{fontenc}
\usepackage[latin1]{inputenc}
\usepackage{babel}
\usepackage{amsmath}
\usepackage{amssymb}
\usepackage{setspace}
\usepackage{esint}
\usepackage{color}
\usepackage{hyperref}
\date{}

\makeatletter

\usepackage{babel}
\usepackage{bbm}
\usepackage{authblk}

\pdfoutput=1

\allowdisplaybreaks
\numberwithin{equation}{section}

\author[1]{Thales Azevedo\thanks{thales@if.ufrj.br}}
\author[2]{Renann Lipinski Jusinskas\thanks{renannlj@fzu.cz}}
\author[3]{Matheus Lize\thanks{matheus.lize@unesp.br}}
\affil[1]{Instituto de F\'isica, Universidade Federal do Rio de Janeiro
\authorcr Av. Athos da Silveira Ramos 149, 21941-972, Rio de Janeiro -- Brazil}
\affil[2]{Institute of Physics of the Czech Academy of Sciences \& CEICO
\authorcr  Na Slovance 2, 182 21, Prague -- Czech Republic}
\affil[3]{Instituto de F\'isica Te\'orica, Universidade Estadual Paulista,
\authorcr Rua Dr. Bento Teobaldo Ferraz 271, 01140-070, S\~ao Paulo -- Brazil}

\makeatother

\begin{document}
\title{\boldmath{\textbf{Bosonic sectorized strings and the $(DF)^{2}$ theory}}}
\maketitle
\begin{abstract}
In this work, we investigate the bosonic chiral string in the sectorized
interpretation, computing its spectrum, kinetic action and $3$-point
amplitudes. As expected, the bosonic ambitwistor string is recovered
in the tensionless limit.

We also consider an extension of the bosonic model with current
algebras. In that case, we compute the effective action and show that
it is essentially the same as the action of the mass-deformed $(DF)^{2}$
theory found by Johansson and Nohle. Aspects which might seem somewhat
contrived in the original construction --- such as the inclusion
of a scalar transforming in some real representation of the gauge
group --- are shown to follow very naturally from the worldsheet
formulation of the theory. \tableofcontents{} 
\end{abstract}

\section{Introduction\label{sec:Introduction}}

\

When Cachazo, He and Yuan (CHY) found their celebrated formulae for
the tree-level scattering amplitudes of massless particles \cite{Cachazo:2013hca,Cachazo:2013iea},
it seemed plausible that those expressions could be obtained from
some worldsheet model. Indeed, it did not take long for Mason and
Skinner to come up with such a model, dubbed ambitwistor strings \cite{Mason:2013sva},
followed by a manifestly supersymmetric version using the pure spinor
formalism \cite{Berkovits:2013xba}. The CHY formulae were later generalized
to different theories \cite{Cachazo:2014nsa,Cachazo:2014xea} and,
again, different ambitwistor strings were proposed as their underlying
worldsheet model \cite{Casali:2015vta}. 

By construction, ambitwistor strings are  two-dimensional chiral theories
which contain no dimensionful parameter. At first, they were thought 
to stem from an infinite tension limit of ordinary string theory, a belief  motivated in part by the fact that the spectrum of
the type II version of the model is identical to that of the corresponding
supergravity. However, for the bosonic and heterotic versions it is
clear that no such procedure is possible, since their spectra do not
match their (super)gravity counterparts.

On the other hand, as observed in \cite{Bandos:2014lja}, ambitwistor
strings are equivalent to the spinor moving frame formulation of the
null superstring --- therefore, tensionless. This idea was supported
by Siegel in \cite{Siegel:2015axg} and other similar results followed
(e.g. \cite{Casali:2016atr}).

It was then noticed that the spectrum of tensionful chiral strings could contain a finite number of massive states \cite{Huang:2016bdd}, depending on the
amount of spacetime supersymmetry. For the type II case, for instance,
the physical spectrum is independent of the string tension. In this
context, the so-called sectorized string model \cite{Jusinskas:2016qjd}
plays an important role. It was introduced as an alternative to the above-mentioned
pure spinor analogue of  ambitwistor strings  \cite{Berkovits:2013xba},
motivated by some  inconsistencies in its heterotic version and difficulties
in coupling it to the $\mathcal{N}=2$ supergravity background \cite{Chandia:2015sfa}.
As such, it was supposed to be a theory for massless particles only.
Nevertheless, it was later shown \cite{Azevedo:2017yjy} that the
heterotic sectorized model actually contains  the $\mathcal{N}=1$
supergravity states together with a single massive multiplet with the same
quantum numbers as the first massive level of the (conventional) open superstring.
This is possible thanks to a dimensionful parameter whose existence
had been overlooked, since the chiral worldsheet action has no parameters.
Moreover, when this parameter is taken to zero, corresponding to a tensionless
limit, one recovers the heterotic ambitwistor string.

In this work, we analyze the bosonic incarnation of the sectorized
model and show how the theory can be interpreted in terms of two 
sectors after a particular gauge-fixing is performed. As in the heterotic
case, the two sectors emulate the left- and right-moving sectors of
the usual string theory, but all worldsheet fields are holomorphic. 
Using methods similar to the ones used in \cite{Berkovits:2018jvm}
for the ambitwistor string --- which in turn were based on \cite{Zwiebach:1992ie}
---, we compute its physical spectrum and the correspondent  kinetic
action. We also analyze the 3-point tree level functions. As expected, the bosonic ambitwistor string is recovered
in the tensionless limit.

We then consider an extension of the bosonic model by including
current algebras. As a main result of this work, the usual methods
are shown to give rise to a worldsheet derivation of the so-called
$(DF)^{2}+{\rm YM}$ theory found by Johansson and Nohle \cite{Johansson:2017srf}.
In particular, the scalar field transforming in some real representation
of the gauge group, whose inclusion might seem somewhat contrived
in the original construction, appears naturally in the sectorized-string
formulation.

Theories whose Lagrangians include a $(DF)^{2}$-type kinetic term
were first introduced as a way of obtaining conformal (super)gravity
amplitudes ($R^2$ gravity, in general) from color-kinematics duality \cite{Bern:2008qj}, and
were shown to admit CHY/ambitwistor representations in \cite{Azevedo:2017lkz}.
Like $R^2$ gravity, such theories contain ``ghost'' states which render them non-unitary. Moreover, in the particular model studied in this paper, a tachyon is also present. It is then natural to ask what physical interest the model might have.

The answer is that scattering amplitudes computed from the $(DF)^{2}$ theories have been recently found
to play a crucial role in the double-copy construction of (ordinary)
bosonic and heterotic string tree-level amplitudes \cite{Azevedo:2018dgo} --- see also \cite{He:2018pol,He:2019drm}. Indeed, just like open superstring amplitudes with external massless states can be expressed in a basis of integrals with coefficients which are nothing but super-Yang--Mills amplitudes \cite{Mafra:2011nv}, the corresponding bosonic open string amplitudes can also be expressed in the same basis, but with  coefficients which come from  $(DF)^{2}+{\rm YM}$. Note that the presence of the tachyon makes perfect sense in this context.

This paper is organized as follows. In section \ref{sec:bosonic-sectorized},
we introduce the sectorized description of the bosonic chiral string,
having the Polyakov action in first-order form as our starting point.
We then investigate the physical spectrum of the model
and analyze its tensionless limit. The kinetic part of its effective action and some results 
on the tree-level three-point amplitudes are also presented. In section \ref{sec:gaugesector},
the bosonic model is extended with the inclusion of current algebras,
and the effective field theory inferred from the three-point functions
is shown to agree with the $(DF)^{2}+{\rm YM}+\phi^{3}$ theory of
Johansson and Nohle. Finally, we present our conclusions and perspectives
in section \ref{sec:conclusion}. The appendix includes further details
on the CFT of current algebras that are relevant for this work.

\section{The bosonic sectorized string\label{sec:bosonic-sectorized}}

In this section we will rederive some known results for chiral bosonic
strings using the sectorized description, including its physical
spectrum and tensionless limit analysis.

\subsection{The Polyakov action in first-order form\label{subsec:Polyakov1st}}

The Polyakov action is given by 
\begin{equation}
S_{P}=\frac{\mathcal{T}}{2}\int d\tau d\sigma\sqrt{-g}\{g^{ij}\partial_{i}X^{m}\partial_{j}X_{m}\},
\end{equation}
where $\mathcal{T}>0$ is the string tension, $g_{ij}$ is the
worldsheet metric (with inverse $g^{ij}$) and $g=\det(g_{ij})$,
with $i,j$ denoting the usual worldsheet coordinates $\tau$ and
$\sigma$. Spacetime indices $m,n,\ldots$ are raised and lowered
with the (mostly plus) Minkowski metric $\eta_{mn}$.

In the first order formulation, one can define a classically equivalent
action, given by 
\begin{multline}
\tilde{S}_{P}=\int d\tau d\sigma\big\{ P_{m}\partial_{\tau}X^{m}-\tfrac{1}{4\mathcal{T}}e_{+}(P_{m}+\mathcal{T}\partial_{\sigma}X_{m})(P^{m}+\mathcal{T}\partial_{\sigma}X^{m})\\
-\tfrac{1}{4\mathcal{T}}e_{-}(P_{m}-\mathcal{T}\partial_{\sigma}X_{m})(P^{m}-\mathcal{T}\partial_{\sigma}X^{m})\big\},\label{eq:Pol1st}
\end{multline}
where $e_{\pm}$ denote the Weyl invariant Lagrange multipliers related
to the worldsheet metric as 
\begin{equation}
e_{\pm}\equiv\frac{1}{g^{\tau\tau}\sqrt{-g}}\mp\frac{g^{\tau\sigma}}{g^{\tau\tau}}.\label{eq:e+-tometric}
\end{equation}

Although not manifestly, the action $\tilde{S}_{P}$ is invariant
under worldsheet reparametrizations, generated by
\begin{equation}
H_{\pm}  \equiv (P_{m}\pm\mathcal{T}\partial_{\sigma}X_{m})(P^{m}\pm\mathcal{T}\partial_{\sigma}X^{m}).\label{eq:ham+-}
\end{equation}The corresponding gauge transformations are given by\begin{subequations}\label{eq:Pol1stgauge}
\begin{eqnarray}
\delta X^{m} & = & \tfrac{1}{2}c_{+}(P^{m}+\mathcal{T}\partial_{\sigma}X^{m})+\tfrac{1}{2}c_{-}(P^{m}-\mathcal{T}\partial_{\sigma}X^{m}),\\
\delta P_{m} & = & \tfrac{\mathcal{T}}{2}\partial_{\sigma}[c_{+}(P^{m}+\mathcal{T}\partial_{\sigma}X^{m})-c_{-}(P^{m}-\mathcal{T}\partial_{\sigma}X^{m})],\nonumber \\
\delta e_{+} & = & \partial_{\tau}c_{+}+c_{+}\partial_{\sigma}e_{+}-e_{+}\partial_{\sigma}c_{+},\\
\delta e_{-} & = & \partial_{\tau}c_{-}-c_{-}\partial_{\sigma}e_{-}+e_{-}\partial_{\sigma}c_{-},
\end{eqnarray}
\end{subequations}where $c_{+}$ and $c_{-}$ are local parameters.

\subsection{The sectorized interpretation}

The quantization of the action \eqref{eq:Pol1st} is straightforward,
and the usual conformal gauge is obtained when we choose $e_{\pm}=1$.
We want to discuss, instead, a particular case of the one-parameter
($\beta$) family of gauges introduced in \cite{Siegel:2015axg},
which can be cast as
\begin{equation}
\begin{array}{cc}
e_{+}=1, & e_{-}={\displaystyle \frac{(1-\beta)}{(1+\beta)}}.\end{array}
\label{eq:siegelgauge}
\end{equation}

For $\beta=0$, the conformal gauge is recovered. We are interested
in the singular gauge $\beta\to\infty$, leading to a chiral
worldsheet action. In this limit, $e_{\pm}=\pm1$. This singular gauge
was proposed in the context of doubled-coordinate field theory in
\cite{Hohm:2013jaa}. After a Wick rotation of the worldsheet coordinate
$\tau$, the gauge-fixed action can be written as
\begin{equation}
S=\frac{1}{2\pi}\int d^{2}z\{P_{m}\bar{\partial}X^{m}+b_{+}\bar{\partial}c_{+}+b_{-}\bar{\partial}c_{-}\},
\label{eq:secbosaction}
\end{equation}
where the gauge parameters $c_{\pm}$ have been promoted to anticommuting
ghosts with corresponding antighosts $b_{\pm}$. All fields in $S$ are holomorphic and the string tension
$\mathcal{T}$ is now hidden.

A few comments about the gauge fixing \eqref{eq:siegelgauge} are in order. For any finite $\beta$, a redefinition of the worldsheet coordinates can always
bring the gauge fixed action to the conformal gauge. This is hardly surprising, since the physical model should be gauge independent.
This was noted by Siegel in \cite{Siegel:2015axg}, but his construction of the
chiral string involved another crucial ingredient related to a change
in the boundary conditions of the action. At any rate,   adopting
the singular gauge ($\beta \to \infty$) is useful since then the delta functions realizing the scattering
equations become explicit.

 It was later noticed that the boundary condition
leading to Siegel\textquoteright s new propagator for the target space
coordinates could in fact be described by the usual string theory
in the conformal gauge ($\beta = 0$), albeit with a different choice of
vacuum \cite{Lee:2017utr}. In the ambitwistor context, this alternative vacuum was investigated in \cite{Casali:2016atr} (and further in \cite{Casali:2017zkz}) and also arises naturally from the quantization
of the action \eqref{eq:secbosaction}. As it turns out, this seems to be the only consistent
vacuum in the singular gauge $\beta\to\infty$. It  might look like
a contradiction, but the key idea here is precisely that this is a
\emph{singular} gauge which effectively leads to a degenerate worldsheet
metric. In other words, the action (2.7) is completely oblivious to
the usual conformal gauge in string theory because this gauge choice
is not invertible (hence, singular). 

In spite of being chiral, the model can be interpreted in terms
of two sectors, namely the ``$+$'' and the ``$-$'', which partially
emulate the left and right movers of the usual bosonic string. Each
sector has its own \emph{characteristic} energy-momentum tensor given
by\begin{subequations} 
\begin{eqnarray}
T_{+} & = & -\frac{1}{4\mathcal{T}}P_{m}^{+}P_{n}^{+}\eta^{mn}-2b_{+}\partial c_{+}+c_{+}\partial b_{+},\\
T_{-} & = & \frac{1}{4\mathcal{T}}P_{m}^{-}P_{n}^{-}\eta^{mn}-2b_{-}\partial c_{-}+c_{-}\partial b_{-},
\end{eqnarray}
\end{subequations}with
\begin{equation}
P_{m}^{\pm}\equiv P_{m}\pm\mathcal{T}\partial X_{m}.\label{eq:Pdef}
\end{equation}
The sectorization is manifest in the BRST charge $Q$:
\begin{eqnarray}
Q & = & Q^{+}+Q^{-},\label{eq:secbosBRST}\\
Q^{\pm} & \equiv & \oint\{c_{\pm}T_{\pm}-b_{\pm}c_{\pm}\partial c_{\pm}\}.
\end{eqnarray}
Nilpotency of $Q$ requires the number of spacetime dimensions to
be $d=26$. 

Note that the complete energy-momentum tensor is given by 
\begin{eqnarray}
T & = & T_{+}+T_{-}\nonumber \\
 & = & -P_{m}\partial X^{m}-b\partial c-\partial(bc)-\tilde{b}\partial\tilde{c}-\partial(\tilde{b}\tilde{c}),
\end{eqnarray}
and it is BRST exact, since $\{Q,(b_{+}+b_{-})\}=T$. In fact, if
we define
\begin{equation}
\begin{array}{ccc}
c\equiv\tfrac{1}{2}(c_{+}+c_{-}), &  & \tilde{c}\equiv\tfrac{1}{2\mathcal{T}}(c_{-}-c_{+}),\\
b\equiv(b_{+}+b_{-}), &  & \tilde{b}\equiv\mathcal{T}(b_{-}-b_{+}),
\end{array}\label{eq:repred}
\end{equation}
the action \eqref{eq:secbosaction} becomes
\begin{equation}
S=\frac{1}{2\pi}\int d^{2}z\{P_{m}\bar{\partial}X^{m}+b\bar{\partial}c+\tilde{b}\bar{\partial}\tilde{c}\},
\end{equation}
while the BRST charge is rewritten as 
\begin{equation}
Q=\oint\{cT-bc\partial c+\tfrac{1}{2}\tilde{c}P^{m}P_{m}+\tfrac{\mathcal{T}^{2}}{2}\tilde{c}(\partial X^{m}\partial X_{m}-2b\partial\tilde{c})\}, \label{BRST01}
\end{equation}
and the familiar Virasoro structure emerges. The tensionless limit
of $Q$ is now very clear: it is precisely the BRST operator introduced
by Mason and Skinner for the bosonic ambitwistor string \cite{Mason:2013sva}.

We will see, however, that the sectorized description is more advantageous
in the cohomology analysis, for it leads to a natural splitting of
the vertex operators in the different mass levels.

\subsection{Physical spectrum\label{subsec:cohomologybos}}

The BRST cohomology at ghost number zero is given by the identity
operator. At ghost number one, the cohomology contains only the zero-momentum
states mapped to the operators $c_{+}P_{m}^{+}$ and $c_{-}P_{m}^{-}$.

Physical states will be defined as elements of the BRST cohomology
with ghost number two and annihilated by the zero mode of $b$. The
latter follows from the usual off-shell condition $(b_{0}-\bar{b}_{0})=0$
on physical states, but adapted to the chiral model. The most general
vertex operator with conformal weight zero satisfying these conditions
can be written as
\begin{equation}
V=V_{0}+V_{+}+V_{-},\label{eq:fullvertex}
\end{equation}
where
\begin{eqnarray}
V_{0} & = & c_{+}c_{-}P_{m}^{+}P_{n}^{-}G^{mn}+\mathcal{T}(c_{+}\partial^{2}c_{+}+c_{-}\partial^{2}c_{-})D+\mathcal{T}(c_{+}\partial^{2}c_{+}-c_{-}\partial^{2}c_{-})E\nonumber \\
 &  & +c_{+}P_{m}^{+}(\partial c_{+}-\partial c_{-})A_{+}^{m}+c_{-}P_{m}^{-}(\partial c_{+}-\partial c_{-})A_{-}^{m},\label{eq:massless}\\
V_{+} & = & c_{+}c_{-}P_{m}^{+}P_{n}^{+}H_{+}^{mn}+c_{-}P_{m}^{+}(\partial c_{+}-\partial c_{-})B_{+}^{m}+c_{+}c_{-}\partial P_{m}^{+}C_{+}^{m}\nonumber \\
 &  & +\mathcal{T}c_{-}\partial^{2}c_{+}F^{+}+b_{+}c_{+}c_{-}(\partial c_{+}-\partial c_{-})G^{+},\label{eq:massive+}\\
V_{-} & = & c_{+}c_{-}P_{m}^{-}P_{n}^{-}H_{-}^{mn}+c_{+}P_{m}^{-}(\partial c_{+}-\partial c_{-})B_{-}^{m}-c_{+}c_{-}\partial P_{m}^{-}C_{-}^{m}\nonumber \\
 &  & +\mathcal{T}c_{+}\partial^{2}c_{-}F^{-}+b_{-}c_{+}c_{-}(\partial c_{+}-\partial c_{-})G^{-}.\label{eq:massive-}
\end{eqnarray}
Here, $G^{mn}$, $H_{\pm}^{mn}$ $A_{\pm}^{m}$, $B_{\pm}^{m}$, $C_{\pm}^{m}$,
$D$, $E$, $F^{\pm}$ and $G^{\pm}$ are the $X$ dependent fields.
This splitting of the terms appearing in the vertex operator is motivated
by their mass-level, as will become clear shortly.

In order to determine the physical degrees of freedom, we will analyze
each of the vertices in \eqref{eq:fullvertex} separately. For $V_{0}$,
the equations of motion imposed by BRST closedness are given by
\begin{equation}
\begin{array}{ccc}
A_{+}^{m}=\tfrac{1}{2}\partial_{n}G^{mn}-\tfrac{1}{2}\partial^{m}(D-E), & \quad & \Box D=\partial_{m}(A_{+}^{m}+A_{-}^{m}),\\
A_{-}^{m}=\tfrac{1}{2}\partial_{n}G^{nm}-\tfrac{1}{2}\partial^{m}(D+E), & \quad & \Box E=\partial_{m}(A_{+}^{m}-A_{-}^{m}),\\
\Box G^{mn}=2\partial^{m}A_{-}^{n}+2\partial^{n}A_{+}^{m}.
\end{array}\label{eq:eomV0}
\end{equation}
These equations become more transparent if we rewrite them in terms of
the fields\begin{subequations}\label{eq:geom}
\begin{eqnarray}
g^{mn} & \equiv & \tfrac{1}{2}(G^{mn}+G^{nm}),\\
b^{mn} & \equiv & \tfrac{1}{2}(G^{mn}-G^{nm}),\\
\phi & \equiv & \tfrac{\mathcal{T}}{2}G^{mn}\eta_{mn}-\mathcal{T}D,\\
g^{m} & \equiv & A_{+}^{m}+A_{-}^{m}-\tfrac{1}{\mathcal{T}}\partial^{m}D,\\
b^{m} & \equiv & A_{+}^{m}-A_{-}^{m}-\tfrac{1}{\mathcal{T}}\partial^{m}E,
\end{eqnarray}
\end{subequations}such that $g^{m}$ and $b^{m}$ have algebraic
solutions, \emph{cf}. \eqref{eq:eomV0},\begin{subequations}
\begin{eqnarray}
g^{m} & = & \partial_{n}g^{mn}-\eta_{np}\partial^{m}g^{np}+\tfrac{2}{\mathcal{T}}\partial^{m}\phi,\\
b^{m} & = & \partial_{n}b^{mn},
\end{eqnarray}
\end{subequations}and\begin{subequations}\label{eq:gbfeom}
\begin{eqnarray}
\Box g^{mn}-\partial_{p}\partial^{n}g^{mp}-\partial_{p}\partial^{m}g^{np}+\eta_{pq}\partial^{m}\partial^{n}g^{pq}-\tfrac{2}{\mathcal{T}}\partial^{m}\partial^{n}\phi & = & 0,\label{eq:graeom}\\
\Box\phi & = & 0,\\
\partial_{p}(\partial^{p}b^{mn}+\partial^{m}b^{np}+\partial^{n}b^{pm}) & = & 0.
\end{eqnarray}
\end{subequations}The gauge transformations, with parameters $\lambda^{m}$
and $\omega^{m}$, are simply
\begin{equation}
\begin{array}{ccccc}
\delta\phi=0, &  & \delta g^{mn}=\partial^{(m}\lambda^{n)}, &  & \delta b^{mn}=\partial^{[m}\omega^{n]}.\end{array}
\end{equation}
It is now easy to identify the field content of the massless sector
described by the vertex $V_{0}$: $\phi$ corresponds to the dilaton,
$b^{mn}$ is the Kalb-Ramond $2$-form and $g^{mn}$ is the graviton,
satisfying the linearized equation of motion \eqref{eq:graeom}.

For the vertices  $V_{+}$ and $V_{-}$, the two sets of equations of motion  are very similar to each other and can be displayed collectively as
\begin{equation}
\begin{array}{ccc}
B_{\pm}^{m}=\partial_{n}H_{\pm}^{mn}-C_{\pm}^{m}-\tfrac{1}{2}\partial^{m}F^{\pm}, &  & \left(\tfrac{1}{4}\Box\mp\mathcal{T}\right)C_{\pm}^{m}=\mathcal{T}B_{\pm}^{m}+\tfrac{1}{2}\partial^{m}G^{\pm},\\
G^{\pm}=\tfrac{\mathcal{T}}{2}H_{\pm}^{mn}\eta_{mn}+\tfrac{1}{2}\partial_{m}C_{\pm}^{m}-\tfrac{3\mathcal{T}}{2}F^{\pm}, &  & \left(\tfrac{1}{4}\Box\mp\mathcal{T}\right)F^{\pm}=\tfrac{1}{2}\partial_{m}B_{\pm}^{m}\mp\tfrac{3}{2}G^{\pm},\\
\left(\tfrac{1}{4}\Box\mp\mathcal{T}\right)H_{\pm}^{mn}=\tfrac{1}{4}\partial^{m}B_{\pm}^{n}+\tfrac{1}{4}\partial^{n}B_{\pm}^{m}\mp\tfrac{1}{4}\eta^{mn}G^{\pm},
\end{array}
\end{equation}
Again, these equations become more transparent after the field redefinitions
\begin{subequations}\label{eq:fieldred_massive}
\begin{eqnarray}
h_{\pm}^{mn} & \equiv & H_{\pm}^{mn}-\tfrac{1}{4\mathcal{T}}(\partial^{n}C_{\pm}^{m}+\partial^{m}C_{\pm}^{n})\pm\tfrac{1}{20\mathcal{T}}(\partial^{m}\partial^{n}\pm\mathcal{T}\eta^{mn})F^{\pm}\nonumber \\
 &  & \mp\tfrac{1}{20\mathcal{T}}(\partial^{m}\partial^{n}\pm\mathcal{T}\eta^{mn})H_{\pm}^{pq}\eta_{pq},\\
f_{\pm} & \equiv & F^{\pm}-H_{\pm}^{mn}\eta_{mn},\\
c_{\pm}^{m} & \equiv & C_{\pm}^{m}\pm\tfrac{1}{10}\partial^{m}H_{\pm}^{np}\eta_{np}\mp\tfrac{1}{10}\partial^{m}F^{\pm},
\end{eqnarray}
\end{subequations}which imply (using $d=26$) that\begin{subequations}\label{eq:h+-eom}
\begin{eqnarray}
\left(\tfrac{1}{4}\Box\mp\mathcal{T}\right)h_{\pm}^{mn} & = & 0,\\
\partial_{n}h_{\pm}^{mn} & = & 0,\\
h_{\pm}^{mn}\eta_{mn} & = & 0,
\end{eqnarray}
\end{subequations}with gauge transformations
\begin{equation}
\begin{array}{ccccc}
\delta h_{\pm}^{mn}=0, &  & \delta f_{\pm}=\pm5\Sigma^{\pm}, &  & \delta c_{\pm}^{m}=\mathcal{T}\Pi_{\pm}^{m}.\end{array}\label{eq:gaugeh+-}
\end{equation}
The fields $f_{\pm}$ and $c_{\pm}^{m}$ are pure gauge, therefore
$h_{\pm}^{mn}$ contain all the physical degrees of freedom, corresponding
to spin $2$ fields with $m^{2}=\pm4\mathcal{T}$.

\subsubsection*{Tensionless limit}

Evidently, in the tensionless limit  all the physical states
are massless. In fact, if we naively take the $\mathcal{T}\to0$ limit
of the vertex \eqref{eq:fullvertex}, it  may seem that Mason and Skinner's
results are recovered \cite{Mason:2013sva}. However, the analysis
of such limit has to be done more carefully precisely because all
the physical states become massless. In other words, the vertices
\eqref{eq:massless}, \eqref{eq:massive+} and \eqref{eq:massive-}
should mix in the tensionless limit. Therefore, we should find a convenient
combination of the fields $G^{mn}$, $H_{\pm}^{mn}$ $A_{\pm}^{m}$,
$B_{\pm}^{m}$, $C_{\pm}^{m}$, $D$, $E$, $F^{\pm}$ and $G^{\pm}$
in \eqref{eq:fullvertex} such that the tensionless limit preserves
the most general form of the vertex operator. The solution is
\begin{eqnarray}
V & = & c\tilde{c}P_{m}P_{n}G_{(1)}^{mn}+c\tilde{c}\partial X_{m}\partial X_{n}G_{(2)}^{mn}+c\tilde{c}P_{m}\partial X_{n}G_{(3)}^{mn}+c\tilde{c}P_{m}\partial X_{n}B^{mn}\nonumber \\
 &  & +c\tilde{c}\partial^{2}X_{m}A_{(1)}^{m}+c\tilde{c}\partial P_{m}A_{(2)}^{m}+\partial\tilde{c}\tilde{c}P_{m}A_{(3)}^{m}+\partial\tilde{c}\tilde{c}\partial X_{m}A_{(4)}^{m}\nonumber \\
 &  & +c\partial\tilde{c}P_{m}A_{(5)}^{m}+c\partial\tilde{c}\partial X_{m}A_{(6)}^{m}+bc\partial\tilde{c}\tilde{c}S_{(1)}+\partial^{2}ccS_{(2)}\nonumber \\
 &  & +\partial^{2}\tilde{c}\tilde{c}S_{(3)}+\partial^{2}c\tilde{c}S_{(4)}+c\partial^{2}\tilde{c}S_{(5)}+\tilde{b}\tilde{c}c\partial\tilde{c}S_{(6)},\label{eq:vertexMN}
\end{eqnarray}
with 
\begin{equation}
\begin{array}{rclcrcl}
G_{(1)}^{mn} & \equiv & 2{\cal{T}}[\tfrac{1}{2}(G^{mn}+G^{nm})+H_{+}^{mn}+H_{-}^{mn}], &  & A_{(5)}^{m} & \equiv & -2{\cal{T}}(A_{+}^{m}+B_{+}^{m}+A_{-}^{m}+B_{-}^{m}),\\
G_{(2)}^{mn} & \equiv & 2\mathcal{T}^{3}[H_{+}^{mn}+H_{-}^{mn}-\tfrac{1}{2}(G^{mn}+G^{nm})], &  & A_{(6)}^{m} & \equiv & -2\mathcal{T}^2(A_{+}^{m}+B_{+}^{m}-A_{-}^{m}-B_{-}^{m}),\\
G_{(3)}^{mn} & \equiv & 4\mathcal{T}^2(H_{+}^{mn}-H_{-}^{mn}), &  & S_{(1)} & \equiv & 2\mathcal{T}^2(G^{+}+G^{-}),\\
B^{mn} & \equiv & -2\mathcal{T}^2(G^{mn}-G^{nm}), &  & S_{(2)} & \equiv & -(2D+F^{+}+F^{-}),\\
A_{(1)}^{m} & \equiv & 2\mathcal{T}^2(C_{+}^{m}+C_{-}^{m}), &  & S_{(3)} & \equiv & -{\mathcal{T}}^{2}(2D-F^{+}-F^{-}),\\
A_{(2)}^{m} & \equiv & 2{\cal{T}}(C_{+}^{m}-C_{-}^{m}), &  & S_{(4)} & \equiv & {\cal{T}}(2E-F^{+}+F^{-}),\\
A_{(3)}^{m} & \equiv & -2\mathcal{T}^2(A_{+}^{m}-B_{+}^{m}-A_{-}^{m}+B_{-}^{m}), &  & S_{(5)} & \equiv & -{\cal{T}}(2E+F^{+}-F^{-}),\\
A_{(4)}^{m} & \equiv & -2\mathcal{T}^{3}(A_{+}^{m}-B_{+}^{m}+A_{-}^{m}-B_{-}^{m}), &  & S_{(6)} & \equiv & 2{\cal{T}}( G^{-}-G^{+}).
\end{array}
\end{equation}

Here the notation  for the fields was chosen so as to agree
with the ambitwistor construction of \cite{Berkovits:2018jvm}, where it was demonstrated that the
free field dynamics associated to the fields above involve higher
derivative operators. This result 
follows naturally from our construction above. For example,
we can show that the equation of motion for the fields $G_{(1)}^{mn}$,
$G_{(2)}^{mn}$ and $G_{(3)}^{mn}$ can be obtained using \eqref{eq:graeom}
and \eqref{eq:h+-eom}, and are given (in the gauge $c^m_\pm=f_\pm=0$) by\begin{subequations}\label{higherders}
\begin{eqnarray}
G_{(2)}^{mn} & = & \tfrac{1}{4}\Box G_{(3)}^{mn}-\mathcal{T}^{2}G_{(1)}^{mn},\\
2G_{(3)}^{mn} & = & \Box G_{(1)}^{mn}-\partial_{p}\partial^{n}G_{(1)}^{mp}-\partial_{p}\partial^{m}G_{(1)}^{np}+\eta_{pq}\partial^{m}\partial^{n}G_{(1)}^{pq}-\tfrac{2}{\mathcal{T}}\partial^{m}\partial^{n}\phi,\label{2G3}\\
\!\!\!\!\!\!(\Box^{2}-16\mathcal{T}^{2})G_{(3)}^{mn} & = & 0,\label{Box2G3}
\end{eqnarray}
\end{subequations}
with $G_{(3)}^{mn}\eta_{mn}=\partial_{n}G_{(3)}^{mn}=0$.

Note that, by substituting \eqref{2G3} into \eqref{Box2G3}, we get an equation involving $\Box^3 G_{(1)}^{mn}$ which, in the tensionless limit, has the same form as the one found in \cite{Berkovits:2018jvm}. Of course, this had to be the case since the vertex operator \eqref{eq:vertexMN} preserves its form as ${\cal T} \to 0$, while the BRST operator reduces to the (bosonic) ambitwistor one, as is evident from \eqref{BRST01}. Indeed, all the other equations of motion can be reproduced in a similar way.

\subsection{Bosonic kinetic action and 3-point amplitudes\label{subsec:bosonic-kinetic}}

As shown above, $G_{(2)}^{mn}$ and $G_{(3)}^{mn}$ can be
seen as auxiliary fields\footnote{\label{nota1}Here the word ``auxiliary'' should not be understood as ``not propagating degrees of freedom,'' but rather that the degrees of freedom represented by these fields can be incorporated in another one which satisfies a higher-derivative equation of motion --- cf. equations \eqref{higherders} above.} which effectively implement a higher derivative
equation of motion for $G_{(1)}^{mn}$. This behavior can be better
understood from another point of view, namely in terms of the effective action
of the model and, in particular, its kinetic part.

Indeed, the kinetic terms associated to $g^{mn}$ and $h_{\pm}^{mn}$
have opposite signs. Physically, this indicates an instability of the
model (ghosts), in agreement with the results of \cite{Huang:2016bdd}.
Such ghosts can usually be  described in terms of higher derivative
theories and this is precisely what happens here.

\subsubsection*{Bosonic kinetic action}

Inspired by Zwiebach's closed string action \cite{Zwiebach:1992ie},
the kinetic action for ambitwistor strings was built in \cite{Berkovits:2018jvm}.
We will use the same prescription for the tensionful model and the
kinetic action will be defined by
\begin{equation}
S=\frac{1}{2}\left\langle V|\partial c\,Q|V\right\rangle 
\end{equation}
where $\left|V\right\rangle $ is the state associated to the vertex
operator \eqref{eq:fullvertex}, obtained from the identity state
$\left|0\right\rangle $ through the state-operator map
\begin{equation}
\left|V\right\rangle =\lim_{z\to0}V(z)\left|0\right\rangle ,
\end{equation}
 and $\left\langle V\right|$ its BPZ conjugate. In order to simplify
the calculations, we will fix the gauge $f_{\pm}=c_{\pm}^{m}=0$ ---
cf. equations \eqref{eq:fieldred_massive} and \eqref{eq:gaugeh+-} ---
and use the auxiliary equations of motion in \eqref{eq:geom} to write
the vertex operator \eqref{eq:fullvertex} in terms of the fields
$g^{mn}$, $b^{mn}$, $\phi$ and $h_{\pm}^{mn}$. 

Now, using the usual ghost measure $\langle c_{\pm}\partial c_{\pm}\partial^{2}c_{\pm}\rangle=2$,
it is straightforward to show that the free action can be cast as
\begin{equation}
S_{bosonic}=S_{0}+S_{+}+S_{-},
\end{equation}
where
\begin{multline}
S_{0}=2\int d^{26}x\{g^{mn}\Box g_{mn}+\partial_{p}g^{mp}\partial^{q}g_{mq}+2(g+\phi)\partial_{m}\partial_{n}g^{mn}\\
-(g+\phi)\Box(g+\phi)+b^{mn}\Box b_{mn}-b^{mn}\partial_{m}\partial^{r}b_{nr}\},
\end{multline}
and
\begin{equation}
S_{\pm}=4\int d^{26}x\{-h_{\pm}^{mn}(\Box\mp4\mathcal{T})h_{\pm mn}+h_{\pm}^{mn}\partial_{n}\partial^{r}h_{\pm mr}-2h_{\pm}\partial_{m}\partial_{n}h_{\pm}^{mn}+h_{\pm}(\Box\mp4\mathcal{T})h_{\pm}\},
\end{equation}
with $g=g^{mn}\eta_{mn}$ and $h_{\pm}=h_{\pm}^{mn}\eta_{mn}$. As
expected, the free field equations of motion derived from $S_{0}$
and $S_{\pm}$ precisely reproduce \eqref{eq:gbfeom} and \eqref{eq:h+-eom}.
The kinetic terms for $g^{mn}$ and $h_{\pm}^{mn}$ have opposite
signs, consistent with the ghost interpretation.

\subsubsection*{3-point amplitudes}

The $3$-point tree level scattering amplitudes for the bosonic chiral string were obtained in \cite{Huang:2016bdd}.
However, it is instructive to redo this analysis here since our unintegrated
vertex operators have a different structure and, in particular, do
not give rise to a Koba--Nielsen factor. For higher point amplitudes, we
would need integrated vertex operators but their definition is still
unknown. 

It will be convenient to gauge fix the vertex operators in \eqref{eq:fullvertex}
and work with momentum eigenstates, such that
\begin{equation}
\begin{array}{cc}
V_{0}=c_{+}c_{-}P_{m}^{+}P_{n}^{-}G^{mn}e^{ik\cdot X}, &  V_{\pm}=c_{+}c_{-}P_{m}^{\pm}P_{n}^{\pm}H_{\pm}^{mn}e^{ik\cdot X},\end{array}
\end{equation}
\noindent where $G^{mn}, H_\pm^{mn}$ are now seen as polarization tensors satisfying $k_{m}G^{mn}=k_{n}G^{mn}=k_{m}H_{\pm}^{mn}=\eta_{mn}H_{\pm}^{mn}=0$.

In order to compute the $3$-point amplitudes, we have to evaluate
its OPE reduction by contracting all $P_{m}^{\pm}$'s with one another 
and with the momentum exponentials $e^{ik\cdot X}$. We also need 
the ghost $3$-point function, which has the usual form
\begin{equation}
\left\langle c_{\pm}(z)c_{\pm}(y)c_{\pm}(w)\right\rangle =(z-y)(y-w)(w-z).\label{eq:3camplitude}
\end{equation}
By virtue of the sectorized description, it is easy to show that the amplitude
factorizes into a product of two open string amplitudes (where $\mathcal{T}\longmapsto-\mathcal{T}$
in the minus sector). With all this in mind, we can compute, for example,
the $3$-point amplitude involving only massless states. The result
is
\begin{equation}
\left\langle V_{0}(z_{1})V_{0}(z_{2})V_{0}(z_{3})\right\rangle =G_{1}^{mn}G_{2}^{pq}G_{3}^{rs}T_{mpr}\bar{T}_{nqs}\delta^{26}(k^{1}+k^{2}+k^{3}),\label{eq:3ptmassless}
\end{equation}
\noindent where
\begin{equation}
T_{mnp}\equiv k_{m}^{2}k_{n}^{3}k_{p}^{1}+2\mathcal{T}(k_{m}^{2}\eta_{np}+k_{n}^{3}\eta_{mp}+k_{p}^{1}\eta_{mn})
\end{equation}
\noindent and $\bar{T}_{nqs}$ is equal to $T_{nqs}$ with the 
sign of $\mathcal{T}$ flipped. The amplitude does not depend on the positions of the vertex operator insertions and is, therefore,  $SL(2,\mathbb{C})$ invariant.
 This result is to some extent expected, since the vertex structure is completely analogous to the ordinary bosonic string and the Koba--Nielsen factors are just $1$ for three massless vertices. However, the $SL(2,\mathbb{C})$ invariance can be shown for \textit{any} $3$-point tree level amplitude, even though the Koba--Nielsen factor is \textit{always} $1$  in the chiral model (there are no contractions between the momentum exponentials since the $XX$ OPE is trivial). The amplitudes factorize in the plus and minus sectors, and there is a precise cancelation of the poles and zeros in $z_{ij}\equiv z_i-z_j$.

\section{Extension of the sectorized model with current algebras\label{sec:gaugesector}}

In this section we will explore the extension of the bosonic sectorized
model in a target space with dimension $d<26$ and the introduction
of current algebras, i.e. a gauge sector. To the action \eqref{eq:secbosaction},
we will add two extra pieces, $S_{C}^{+}$ and $S_{C}^{-}$, describing
two current algebras. The new BRST charge preserves its form in \eqref{eq:secbosBRST}
but now with 
\begin{eqnarray}
T_{+} & = & -\tfrac{1}{4\mathcal{T}}P_{m}^{+}P_{n}^{+}\eta^{mn}-2b_{+}\partial c_{+}+c_{+}\partial b_{+}+T_{C}^{+},\\
T_{-} & = & \tfrac{1}{4\mathcal{T}}P_{m}^{-}P_{n}^{-}\eta^{mn}-2b_{-}\partial c_{-}+c_{-}\partial b_{-}+T_{C}^{-},
\end{eqnarray}
where $T_{C}^{\pm}$ denotes the energy-momentum tensor associated
to different group manifolds with central charge 
\begin{equation}
c^{(\pm)}=26-d.\label{eq:centralchargecurrent}
\end{equation}

For now we will focus on the ``$-$'' sector, which contains the
tachyonic excitations. The inclusion of the ``$+$'' sector, which
has an analogous structure, will be discussed in subsection \ref{subsec:+gaugesector}.

Let us consider an affine Lie algebra associated to some group $G$,
with structure constants $f_{ab}^{\phantom{ab}c}$ ($a,b,\ldots=1$
to $\dim G$) and level $k$. The addition of $S_{C}^{-}$ to the
action allows us to define currents $J_{a}$ which are primary conformal
fields and satisfy the OPE 
\begin{equation}
J_{a}(z)\,J_{b}(y)\sim\frac{k\delta_{ab}}{(z-y)^{2}}+if_{ab}^{\hphantom{ab}c}\frac{J_{c}(y)}{(z-y)}.\label{eq:OPEJ-J}
\end{equation}
Here the group generators have been orthonormalized such that the
metric $\delta_{ab}$ corresponds to a Kronecker delta, and we will
make no further distinction between upper and lower indices.

The energy-momentum tensor of the algebra can be obtained using the
Sugawara construction and is given by 
\begin{equation}
T_{C}^{-}\equiv\frac{1}{2(k+g)}\left(J_{a},J_{a}\right),
\end{equation}
where $g$ is the dual Coxeter number, defined through 
\begin{equation}
f_{acd}f_{bcd}=2g\delta_{ab}.
\end{equation}
We use the ordering prescription 
\begin{equation}
\left(A,B\right)(y)\equiv\tfrac{1}{2\pi i}\oint\frac{dz}{(z-y)}A(z)B(y),\label{eq:ordering}
\end{equation}
which can be understood as the product of two operators $A(z)$ and
$B(y)$ in the limit $z\to y$, with singular terms removed.

It is then straightforward to compute the central charge of this model,
which is given by
\begin{eqnarray}
c^{(-)} & = & \frac{k\Delta}{(k+g)},\nonumber \\
 & \stackrel{!}{=} & 26-d,\label{eq:centralcharge}
\end{eqnarray}
where
\begin{equation}
\Delta\equiv\delta^{ab}\delta_{ab}=\dim G.
\end{equation}
The second equality in \eqref{eq:centralcharge} comes from imposing the nilpotency of the BRST
operator and constrains the group $G$ and the level $k$ of the current
algebra. For example, for a target space with $d=10$ one of the solutions
is $G=SO(32)$ and $k=1$, while for $d=4$ we can have $G=SU(5)$
and $k=55$, and so on. Further constraints on the group should arise
from the analysis of anomalies but this will not be discussed in this
work.

\subsection{Physical spectrum\label{subsec:physicalspectrum}}

The BRST cohomology now includes additional states with corresponding
vertex operators containing the currents $J_{a}$, expressed as 
\begin{eqnarray}
V_{J} & = & c_{+}c_{-}P_{m}^{+}J_{a}F_{a}^{m}+c_{-}(\partial c_{+}-\partial c_{-})J_{a}F^{a}+c_{+}c_{-}\partial J_{a}S_{a}\nonumber \\
 &  & +c_{+}c_{-}P_{m}^{-}J_{a}G_{a}^{m}+c_{+}(\partial c_{+}-\partial c_{-})J_{a}G^{a}+c_{+}c_{-}J_{\alpha}\varphi_{\alpha}.\label{eq:vertexJ}
\end{eqnarray}
Here $F_{a}^{m}$, $G_{a}^{m}$, $S^{a}$, $F_{a}$, $G_{a}$ and
$\varphi_{\alpha}$ are target space fields. The index $\alpha$ belongs
to a traceless-symmetric bi-adjoint representation of the group $G$ (see appendix),
with dimension 
\begin{equation}
\Delta(\alpha)=\frac{\Delta(\Delta+1)}{2}-1.
\end{equation}
$J_{\alpha}$ is a primary conformal weight $2$ operator defined
as 
\begin{equation}
J_{\alpha}\equiv\left(C^{-1}\right)_{\alpha ab}J_{(ab)},\label{eq:Jalpha}
\end{equation}
where $J_{(ab)}$ is given by the traceless-symmetric ordered product
of two currents, \emph{i.e.} 
\begin{equation}
J_{(ab)}\equiv\tfrac{1}{2}\left(J_{a},J_{b}\right)+\tfrac{1}{2}\left(J_{b},J_{a}\right)-\tfrac{2(k+g)}{\Delta}\delta_{ab}T_{C}^{-},
\end{equation}
and $\left(C^{-1}\right)_{\alpha ab}$ are the inverse of the Clebsch-Gordan
coefficients, $C_{\alpha ab}$. The properties of these coefficients
will be discussed in the next subsection and in the appendix. Observe that we could have
considered also the trace contribution in the vertex, \emph{e.g.}
$c_{+}c_{-}T_{C}^{-}\varphi$. However, the field $\varphi$ couples
only to the vertex $V_{-}$ in subsection \eqref{subsec:cohomologybos}
and does not change the physical content of the model.

The BRST invariance of the vertex $V_{J}$ implies the following equations
of motion\begin{subequations}\label{eq:eomJ} 
\begin{eqnarray}
(\Box+4\mathcal{T})\varphi_{\alpha} & = & 0,\\
F_{a} & = & \tfrac{1}{2}\partial_{m}F_{a}^{m},\label{eq:Fsol}\\
G_{a} & = & \tfrac{1}{2}\partial_{m}G_{a}^{m}-S_{a},\label{eq:Gsol}\\
\partial_{n}(\partial^{m}F_{a}^{n}-\partial^{n}F_{a}^{m}) & = & 0,\\
\partial_{n}(\partial^{m}G_{a}^{n}-\partial^{n}G_{a}^{m}) & = & 4\mathcal{T}G_{a}^{m}+2\partial^{m}S_{a},
\end{eqnarray}
\end{subequations}and the gauge transformations can be summarized
as
\begin{equation}
\begin{array}{ccccc}
\delta F_{a}^{m}=\partial^{m}\Lambda_{a}, &  & \delta G_{a}^{m}=\partial^{m}\Omega_{a}, &  & \delta S_{a}=-2\mathcal{T}\Omega_{a}.\end{array}\label{eq:gaugeJ}
\end{equation}
Since $S_{a}$ is pure gauge, the physical states described by the
vertex \eqref{eq:vertexJ} correspond to a massless vector $F_{a}^{m}$
and two fields with negative mass-squared $m^{2}=-4\mathcal{T}$ namely the scalar $\varphi_{\alpha}$
and the vector $G_{a}^{m}$.

In parallel to subsection \ref{subsec:cohomologybos}, we can prepare
the vertex $V_{J}$ for the tensionless limit analysis. Considering
the redefinitions of the worldsheet ghosts of \eqref{eq:repred},
$V_{J}$ can be rewritten as 
\begin{equation}
\tfrac{1}{2\mathcal{T}}V_{J}=c\tilde{c}J_{\alpha}\varphi_{\alpha}+c\tilde{c}P_{m}J_{a}A_{a}^{m}+c\tilde{c}\partial X_{m}J_{a}B_{a}^{m}-c\partial\tilde{c}J_{a}A_{a}-\tilde{c}\partial\tilde{c}J_{a}B_{a}.\label{eq:gaugeTprepared}
\end{equation}
Here, the fields $A_{a}$, $A_{a}^{m}$, $B_{a}$ and $B_{a}^{m}$
are defined in terms of $F_{a}^{m}$, $G_{a}^{m}$, $F_{a}$ and $G_{a}$
as
\begin{equation}
\begin{array}{ccc}
A_{a}\equiv F_{a}+G_{a}, &  & B_{a}\equiv\mathcal{T}(F_{a}-G_{a}),\\
A_{a}^{m}\equiv F_{a}^{m}+G_{a}^{m}, &  & B_{a}^{m}\equiv\mathcal{T}(F_{a}^{m}-G_{a}^{m}),
\end{array}\label{eq:fieldredgauge}
\end{equation}
with gauge transformations $\delta A_{a}^{m}=\partial^{m}\Lambda_{a}$
and $\delta B_{a}^{m}=\mathcal{T}\partial^{m}\Lambda_{a}$.

Their equations of motion follow from \eqref{eq:eomJ} and are given
by
\begin{equation}
\begin{array}{ccc}
A_{a}=\tfrac{1}{2}\partial_{m}A_{a}^{m}, &  & \mathcal{T}A_{a}^{m}-\tfrac{1}{2}\partial_{n}F_{a}^{mn}=B_{a}^{m},\\
B_{a}=\tfrac{1}{2}\partial_{m}B_{a}^{m}, &  & (\Box+4\mathcal{T})\partial_{n}F_{a}^{mn}=0.
\end{array}\label{eq:eomABs}
\end{equation}
Therefore, the physical spectrum can be described in terms of only
two fields, $\varphi_{\alpha}$ and $A_{a}^{m}$. The vector $B_{a}^{m}$
is auxiliary, helping to implement a quartic equation of motion for
$A_{a}^{m}$, which carries the degrees of freedom of both the massless
and the massive vector fields, $F_{a}^{m}$ and $G_{a}^{m}$. Note,
in particular, the tensionless limit renders a massless spectrum
with equations of motion $\Box\varphi_{\alpha}=\Box^{2}A_{a}^{m}=0$.
As in the bosonic model of section \ref{sec:bosonic-sectorized},
this behavior can be easily observed when analyzing the effective
field theory associated to the model, which will be done in subsection
\ref{subsec:EFT}. The first step will be to determine the $3$-point
amplitudes using the vertex \eqref{eq:vertexJ}.

\subsection{3-point amplitudes}

In order to compute the $3$-point amplitude 
\begin{equation}
\mathcal{A}_{3}\equiv\left\langle V_{J}(z)V_{J}(y)V_{J}(w)\right\rangle ,\label{eq:3ptVJ}
\end{equation}
we need to provide further details on the current algebra CFT, in
particular the OPE's involving the operator $J_{\alpha}$ defined
in \eqref{eq:Jalpha} and the properties of the Clebsch-Gordan coefficients
$C_{\alpha ab}$.

The operator $J_{\alpha}$ satisfies the following OPE's:\begin{subequations}\label{eq:OPEscurrents}
\begin{eqnarray}
T_{C}^{-}(z)\,J_{\alpha}(y) & \sim & \frac{2J_{\alpha}}{(z-y)^{2}}+\frac{\partial J_{\alpha}}{(z-y)},\\
J_{a}(z)\,J_{\alpha}(y) & \sim & C_{\alpha ab}\frac{J_{b}}{(z-y)^{2}}-\left(T_{a}\right)_{\alpha\beta}\frac{J_{\beta}}{(z-y)},\label{eq:OPEJ-JJ}\\
J_{\alpha}(z)\,J_{\beta}(y) & \sim & \frac{k\delta_{\alpha\beta}}{(z-y)^{4}}-\left(T_{a}\right)_{\alpha\beta}\left\{ \frac{J_{a}}{(z-y)^{3}}+\tfrac{1}{2}\frac{\partial J_{a}}{(z-y)^{2}}+\tfrac{1}{6}\frac{\partial^{2}J_{a}}{(z-y)}\right\} \nonumber \\
 &  & +d_{\alpha\beta\gamma}\left\{ \frac{J_{\gamma}}{(z-y)^{2}}+\tfrac{1}{2}\frac{\partial J_{\gamma}}{(z-y)}\right\} \nonumber \\
 &  & +d_{\alpha\beta abc}\frac{J_{(abc)}}{(z-y)}+d_{\alpha\beta[ab]}\frac{J_{[ab]}}{(z-y)}+e_{a\alpha\beta}\frac{\left(J_{a},T_{C}^{-}\right)}{(z-y)}.\label{eq:OPEJJ-JJ}
\end{eqnarray}
\end{subequations}The first OPE states that $J_{\alpha}$ is a primary
operator of conformal dimension $2$. The second OPE is connected
to the definition of the Clebsch-Gordan coefficients (quadratic pole)
and the group transformation of $J_{\alpha}$ (simple pole). $\left(T_{a}\right)_{\alpha\beta}$
denotes the group generators in the traceless bi-adjoint representation
of the group $G$ and satisfy\begin{subequations} 
\begin{eqnarray}
[T_{a},T_{b}]_{\alpha\beta} & = & if_{abc}\left(T_{c}\right)_{\alpha\beta},\\
\left(T_{a}\right)_{\alpha\beta} & \equiv & 2if_{abc}C_{\alpha(ce)}\left(C^{-1}\right)_{\beta(be)},\\
\left(T_{a}T_{b}\right)_{\alpha\alpha} & = & 2g(\Delta+2)\delta_{ab},\\
\left(T_{a}T_{a}\right)_{\alpha\beta} & = & 4g\delta_{\alpha\beta}-2f_{abc}f_{ade}C_{\alpha(ce)}\left(C^{-1}\right)_{\beta(bd)},
\end{eqnarray}
\end{subequations}

The OPE \eqref{eq:OPEJJ-JJ} can be used to define
the $2$-point and $3$-point functions involving only $J_{\alpha}$'s.
Operators of conformal dimension $3$ appear in the last line (with
numerical coefficients $d_{\alpha\beta abc}$, $d_{\alpha\beta[ab]}$
and $e_{a\alpha\beta}$) but they do not contribute to $\mathcal{A}_{3}$.
$J_{(abc)}$ is the totally symmetric traceless normal ordered product
of $J_{a}$, $J_{b}$ and $J_{c}$, and $J_{[ab]}$ is the antisymmetric
product $(J_{a},J_{b})-(J_{b},J_{a})$.

The Clebsch-Gordan coefficients $C_{\alpha ab}$ are defined in such
a way that\begin{subequations} 
\begin{eqnarray}
C_{\alpha ab}\left(C^{-1}\right)_{\beta ab} & = & \delta_{\alpha\beta},\\
C_{\alpha ab}\left(C^{-1}\right)_{\alpha cd} & = & \delta_{(ab)(cd)},\\
C_{\alpha ab}C_{\alpha cd} & = & \Delta_{(ab)(cd)}+2k\delta_{(ab)(cd)},\\
C_{\alpha ab}C_{\beta ab} & = & f_{ade}f_{bce}C_{\beta ab}\left(C^{-1}\right)_{\alpha cd}+2k\delta_{\alpha\beta},
\end{eqnarray}
\end{subequations}with\begin{subequations} 
\begin{eqnarray}
\delta_{(ab)(cd)} & \equiv & \tfrac{1}{2}\delta_{ac}\delta_{bd}+\tfrac{1}{2}\delta_{ad}\delta_{bc}-\tfrac{1}{\Delta}\delta_{ab}\delta_{cd},\\
\Delta_{(ab)(cd)} & \equiv & \tfrac{1}{2}f_{ade}f_{bce}+\tfrac{1}{2}f_{ace}f_{bde}-\tfrac{2g}{\Delta}\delta_{ab}\delta_{cd}.
\end{eqnarray}
\end{subequations}Finally, the coefficient $d_{\alpha\beta\gamma}$
is defined as 
\begin{equation}
d_{\alpha\beta\gamma}\equiv\left(C^{-1}\right)_{\beta ab}\left[\left(T_{a}T_{b}\right)_{\alpha\gamma}+2C_{\alpha ac}C_{\gamma bc}\right],
\end{equation}
or 
\begin{equation}
C_{\beta ab}d_{\alpha\beta\gamma}=\tfrac{1}{2}\left(T_{a}T_{b}\right)_{\alpha\gamma}+C_{\alpha ae}C_{\gamma be}+\left(a\leftrightarrow b\right)-\textrm{trace}.
\end{equation}
Although not manifestly, $d_{\alpha\beta\gamma}$ is traceless, \emph{i.e.}
$d_{\alpha\alpha\gamma}=0$, and completely symmetric in the exchange
of any pair of indices.

The $2$-point amplitudes involving the gauge currents can be easily
determined through the OPE's \eqref{eq:OPEJ-J}, \eqref{eq:OPEJ-JJ}
and \eqref{eq:OPEJJ-JJ}, and are given by\begin{subequations} 
\begin{eqnarray}
\left\langle J_{a}(z)J_{b}(y)\right\rangle  & = & \frac{k\delta_{ab}}{(z-y)^{2}},\\
\left\langle J_{a}(z)J_{\alpha}(y)\right\rangle  & = & 0,\\
\left\langle J_{\alpha}(z)J_{\beta}(y)\right\rangle  & = & \frac{k\delta_{\alpha\beta}}{(z-y)^{4}}.
\end{eqnarray}
\end{subequations} The $3$-point amplitudes are now straightforward
to compute. They can be summarized as\begin{subequations} 
\begin{eqnarray}
\left\langle J_{a}(z)J_{b}(y)J_{c}(w)\vphantom{J_{\gamma}}\right\rangle  & = & -ikf_{abc}(z-y)^{-1}(y-w)^{-1}(w-z)^{-1},\\
\left\langle J_{\alpha}(z)J_{a}(y)J_{b}(w)\vphantom{J_{\gamma}}\right\rangle  & = & kC_{\alpha ab}(z-y)^{-2}(w-z)^{-2},\\
\left\langle J_{\alpha}(z)J_{\beta}(y)J_{a}(w)\vphantom{J_{\gamma}}\right\rangle  & = & k\left(T_{a}\right)_{\alpha\beta}(z-y)^{-3}(y-w)^{-1}(w-z)^{-1},\\
\left\langle J_{\alpha}(z)J_{\beta}(y)J_{\gamma}(w)\right\rangle  & = & kd_{\alpha\beta\gamma}(z-y)^{-2}(y-w)^{-2}(w-z)^{-2}.
\end{eqnarray}
\end{subequations}

As one last step before evaluating \eqref{eq:3ptVJ}, it will be convenient
to fix the gauge degrees of freedom of $V_{J}$. Using the gauge transformations
\eqref{eq:gaugeJ}, we will choose $S_{a}=0$. In this gauge, $\partial_{m}G_{a}^{m}=0$
as a consequence of the equations of motion. We can use the remaining
parameter to fix the transversal gauge for the massless vector, such
that the vertex is simplified to 
\begin{equation}
V_{J}=c_{+}c_{-}P_{m}^{+}J_{a}F_{a}^{m}+c_{+}c_{-}P_{m}^{-}J_{a}G_{a}^{m}+c_{+}c_{-}J_{\alpha}\varphi_{\alpha}.
\end{equation}

Using the tree level measure for the ghosts \eqref{eq:3camplitude},
the $3$-point amplitude \eqref{eq:3ptVJ} can be computed to be 
\begin{eqnarray}
\mathcal{A}_{3} & = & kd_{\alpha\beta\gamma}\left\langle \varphi_{\alpha}\varphi_{\beta}\varphi_{\gamma}\right\rangle -3k\left(T_{a}\right)_{\alpha\beta}\left\langle \varphi_{\alpha}\partial_{m}\varphi_{\beta}(F_{a}^{m}+G_{a}^{m})\right\rangle \nonumber \\
 &  & -3kC_{\alpha ab}\left\langle \partial_{m}\partial_{n}\varphi_{\alpha}(F_{a}^{m}+G_{a}^{m})(F_{b}^{n}+G_{b}^{n})\right\rangle \nonumber \\
 &  & -ikf_{abc}\left\langle \partial_{p}(F_{a}^{m}+G_{a}^{m})\partial_{m}(F_{b}^{n}+G_{b}^{n})\partial_{n}(F_{c}^{p}+G_{c}^{p})\right\rangle \nonumber \\
 &  & +6k\mathcal{T}C_{\alpha ab}\eta_{mn}\left\langle \varphi_{\alpha}(F_{a}^{m}-G_{a}^{m})(F_{b}^{n}+G_{b}^{n})\right\rangle \nonumber \\
 &  & -6ik\mathcal{T}f_{abc}\eta_{mn}\left\langle (F_{a}^{m}-G_{a}^{m})\partial_{p}(F_{b}^{n}+G_{b}^{n})(F_{c}^{p}+G_{c}^{p})\right\rangle .\label{eq:3pointamplitudegauge}
\end{eqnarray}
Observe that $\mathcal{A}_{3}$ is at most linear in $\mathcal{T}(F_{a}^{m}-G_{a}^{m})$.
If we look at the vertex \eqref{eq:gaugeTprepared}, this is easy
to understand because the $3$-point amplitudes with two or three
$B_{a}^{m}$'s vanish trivially.

In principle, $4$-point amplitudes can be computed using the results
of Siegel in \cite{Siegel:2015axg}. Currently, however, there is
no clear definition of the integrated vertex operators and higher
point amplitudes cannot be \emph{directly} computed from the chiral
model. This problem will be dealt with in a separate paper by one
of the authors.

In the next subsection we will propose an effective field theory action
for the field content of the previous subsection.

\subsection{Effective field theory: $(DF)^{2}+\rm{YM}$\label{subsec:EFT}}

As the main result of this paper, we would like to argue that the
effective field theory action corresponding to this extension of the
bosonic sectorized model is precisely the action of the $(DF)^{2}+{\rm YM}$
theory constructed in \cite{Johansson:2017srf}. Indeed, we have already
shown the spectrum to be the same. The action can be decomposed as
\begin{equation}
S_{eff}=S_{J}^{0}+S_{J}^{int},
\end{equation}
where $S_{J}^{0}$ is the kinetic part of the action and $S_{J}^{int}$
corresponds to the interactions.

For the kinetic part, we will proceed like in subsection \eqref{subsec:bosonic-kinetic}.
For the interaction part, we will analyze the possible vertices that
give rise to the $3$-point amplitudes displayed in \eqref{eq:3pointamplitudegauge}.
Next, we will require the non-linear gauge invariance of the resulting
model in order to finally propose its effective action.

\subsubsection{Kinetic action}

As stated above, we will define the kinetic action as 
\begin{equation}
S_{J}^{0}\equiv\left\langle V_{J}\left|\partial c\,Q\right|V_{J}\right\rangle ,
\end{equation}
up to normalization. 

In order to further simplify the computation, we will consider the
algebraic solutions \eqref{eq:Fsol} and \eqref{eq:Gsol}, such that
\begin{eqnarray}
\partial c[Q,V_{J}] & = & \tfrac{1}{4\mathcal{T}}c_{+}c_{-}\partial c_{+}\partial c_{-}J_{a}P_{m}^{+}[\partial_{n}(\partial^{n}F_{a}^{m}-\partial^{m}F_{a}^{n})]\nonumber \\
 &  & +\tfrac{1}{4\mathcal{T}}c_{+}c_{-}\partial c_{+}\partial c_{-}J_{a}P_{m}^{-}[\partial_{n}(\partial^{n}G_{a}^{m}-\partial^{m}G_{a}^{n})+4\mathcal{T}G_{a}^{m}]\nonumber \\
 &  & +\tfrac{1}{4\mathcal{T}}c_{+}c_{-}\partial c_{+}\partial c_{-}\partial J_{a}[2\mathcal{T}\partial_{m}G_{a}^{m}]\nonumber \\
 &  & +\tfrac{1}{4\mathcal{T}}c_{+}c_{-}\partial c_{+}\partial c_{-}J_{\alpha}[\Box\varphi_{\alpha}+4\mathcal{T}\varphi_{\alpha}].
\end{eqnarray}
It is then straightforward to show that 
\begin{multline}
S_{J}^{0}=\int d^{d}x\{\varphi_{\alpha}(\Box\varphi_{\alpha}+4\mathcal{T}\varphi_{\alpha})-2\mathcal{T}F_{ma}(\Box F_{a}^{m}-\partial^{m}\partial_{n}F_{a}^{n})\\
+2\mathcal{T}G_{ma}(\Box G_{a}^{m}+4\mathcal{T}G_{a}^{m}-\partial^{m}\partial_{n}G_{a}^{n})\}.
\end{multline}

Note that the kinetic terms of the fields $F_{a}^{m}$ and $G_{a}^{m}$
have opposite sign in $S_{J}^{0}$. Technically, the sign difference
can be traced back to the OPE's of $P_{m}^{+}$ and $P_{m}^{-}$ with
themselves. As discussed previously, this indicates an instability
of the model and we can again reinterpret it in terms of a higher
derivative theory. In fact, as we will now show, this behavior is
more transparent if we rewrite the action in terms of the vectors
$A_{a}^{m}$ and $B_{a}^{m}$ defined in \eqref{eq:fieldredgauge}.
The kinetic action can then be cast as 
\begin{equation}
S_{J}^{0}=\int d^{d}X\,\{\varphi_{\alpha}(\Box\varphi_{\alpha}+4\mathcal{T}\varphi_{\alpha})+2B_{ma}\partial_{n}F_{a}^{mn}+2(B_{a}^{m}-\mathcal{T}A_{a}^{m})(B_{ma}-\mathcal{T}A_{ma})\},
\end{equation}
with 
\begin{equation}
F_{a}^{mn}\equiv\partial^{m}A_{a}^{n}-\partial^{n}A_{a}^{m}.
\end{equation}

Ignoring for now the interaction terms, observe that the equation
of motion for $B_{a}^{m}$ is algebraic, given by 
\begin{equation}
B_{a}^{m}=\mathcal{T}A_{a}^{m}+\tfrac{1}{2}\partial_{n}F_{a}^{nm}.\label{eq:eomBlivre}
\end{equation}
If we replace this solution back in the action, we obtain 
\begin{equation}
S_{J}^{0}|_{B}=\int d^{d}X\,\{\varphi_{\alpha}(\Box\varphi_{\alpha}+4\mathcal{T}\varphi_{\alpha})+\mathcal{T}F_{a}^{mn}F_{mna}-\tfrac{1}{2}\partial_{n}F_{a}^{mn}\partial^{p}F_{mpa}\}.\label{eq:S_J^0}
\end{equation}
This action can be identified with the kinetic part of the $(DF)^{2}+{\rm YM}$
theory constructed in \cite{Johansson:2017srf}. Note that the propagator
of $A_{a}^{m}$ is given in momentum space by 
\begin{equation}
G_{ab}^{mn}(p)=\frac{i\eta^{mn}\delta_{ab}}{p^{2}(p^{2}-4{\cal T})}.
\end{equation}
The pole structure of this propagator agrees with the interpretation
given after equation \eqref{eq:eomABs} that $A_{a}^{m}$ effectively
describes the massless and the massive vector fields, $F_{a}^{m}$
and $G_{a}^{m}$.

\subsubsection{Cubic vertices and the effective action}

As it turns out, the procedure of integrating $B_{a}^{m}$ out can
be partially extended to interactions. We say ``partially'' because
in this paper we consider only unintegrated vertex operators, therefore
only $3$-point tree level amplitudes. We expect this integration
to hold for higher point vertices as well.

By looking at $\mathcal{A}_{3}$ in \eqref{eq:3pointamplitudegauge},
it is easy to show that the $3$-point vertices in terms of the vectors
$A_{a}^{m}$ and $B_{a}^{m}$ can be schematically expressed as 
\begin{equation}
\begin{array}{rclcrcl}
\varphi^{3} & \sim & d_{\alpha\beta\gamma}\varphi_{\alpha}\varphi_{\beta}\varphi_{\gamma}, &  & \varphi A^{2} & \sim & C_{\alpha ab}\varphi_{\alpha}\partial_{n}A_{a}^{m}\partial_{m}A_{b}^{n},\\
\varphi^{2}A & \sim & \left(T_{a}\right)_{\alpha\beta}\varphi_{\alpha}\partial_{m}\varphi_{\beta}A_{a}^{m}, &  & A^{3} & \sim & if_{abc}\partial_{p}A_{a}^{m}\partial_{m}A_{b}^{n}\partial_{n}A_{c}^{p},\\
\varphi AB & \sim & C_{\alpha ab}\eta_{mn}\varphi_{\alpha}B_{a}^{m}A_{b}^{n}, &  & A^{2}B & \sim & if_{abc}\eta_{mn}B_{a}^{m}\partial_{p}A_{b}^{n}A_{c}^{p}.
\end{array}
\end{equation}

The idea now is to analyze the possible gauge invariant interactions
that can generate these vertices after integrating out $B_{a}^{m}$,
which is at most linear in the expressions above. The equation of
motion for $B_{a}^{m}$ in \eqref{eq:eomBlivre} gets modified to
\begin{equation}
B_{a}^{m}=\mathcal{T}A_{a}^{m}+\tfrac{1}{2}\partial_{n}F_{a}^{nm}+c_{\#}C_{\alpha ab}\varphi_{\alpha}A_{mb}+id_{\#}f_{abc}\eta_{mn}\partial_{p}A_{b}^{n}A_{c}^{p}+\ldots,
\end{equation}
where $c_{\#}$ and $d_{\#}$ are numerical constants and the dots
contain other terms necessary to generate the correct gauge transformation
for $B_{a}^{m}$ (remember that the onshell $3$-point amplitude $\mathcal{A}_{3}$
was computed using gauge-fixed vertex operators). Taking this into
consideration and replacing $B_{a}^{m}$ in the action, we can show
that \emph{all} $3$-point vertices come from the operators
\[
\begin{array}{cccccc}
C_{\alpha ab}\varphi_{\alpha}F_{a}^{mn}F_{mnb}, & (D\varphi)^{2}, & (DF)^{2}, & F^{3}, & F^{2}, & d_{\alpha\beta\gamma}\varphi_{\alpha}\varphi_{\beta}\varphi_{\gamma},\end{array}
\]
where $F_{a}^{mn}$ was redefined to be the non-Abelian field strength
\begin{equation}
F_{a}^{mn}\equiv(\partial^{m}A_{a}^{n}-\partial^{n}A_{a}^{m})+igf_{abc}A_{b}^{m}A_{c}^{n},
\end{equation}
with coupling constant $g$, and $D^{m}$ denotes the covariant derivative
with respect to the vector $A_{a}^{m}$. The form of the higher point
vertices ($4$, $5$ and $6$) is severely restricted by the non-linear
gauge invariance of the effective action. Some contributions naturally
appear after integrating out $B_{a}^{m}$ and we expect them to combine
with the input coming from higher-point amplitudes, which involve
integrated vertex operators.

Finally, we propose the effective field theory action of the model
to be 
\begin{multline}
S_{eff}=\int d^{d}x\Big\{\tfrac{1}{2}(D_{n}F_{a}^{mn})^{2}-\mathcal{T}F_{a}^{mn}F_{mna}+\tfrac{1}{2}D_{m}\varphi_{\alpha}D^{m}\varphi_{\alpha}-2\mathcal{T}(\varphi^{\alpha})^{2}\\
+\tfrac{g}{3}f_{abc}F_{\hphantom{m}na}^{m}F_{\hphantom{n}pb}^{n}F_{\hphantom{p}mc}^{p}+\tfrac{g}{2}C_{\alpha ab}\varphi_{\alpha}F_{a}^{mn}F_{mnb}+\tfrac{g}{3!}d_{\alpha\beta\gamma}\varphi_{\alpha}\varphi_{\beta}\varphi_{\gamma}\Big\},\label{eq:Effectiveaction-minus}
\end{multline}
where $g$ is the coupling constant. This action describes the $(DF)^{2}+{\rm YM}$
theory of \cite{Johansson:2017srf}.

Moreover, if we include the ``$+$'' sector mentioned in the beginning
of this section, the effective field theory action describes a more
general model with a mirrored set of fields. In particular, if we
restrict the gauge symmetry of the ``$+$'' sector to be instead
a global symmetry, the effective action describes the $(DF)^{2}+\text{YM}+\phi^{3}$
theory. This will be shown next.

\subsection{Including the other gauge sector: $(DF)^{2}+\rm{YM}+\phi^{3}$\label{subsec:+gaugesector}}

We will consider for the ``$+$'' sector an affine Lie algebra associated
to a group $\hat{G}$ (with structure constants $\hat{f}_{AB}^{\hphantom{AB}C}$)
and level $\hat{k}$. Apart from the central charge constraint \eqref{eq:centralchargecurrent},
$\{\hat{G},\hat{k}\}$ are independent of $\{G,k\}$, from the ``$-$''
sector. The new currents, $\hat{J}_{A}$, are completely analogous
to the ones discussed there, \emph{e.g.} they satisfy the OPE 
\begin{equation}
\hat{J}_{A}(z)\,\hat{J}_{B}(y)\sim\frac{\hat{k}\delta_{AB}}{(z-y)^{2}}+i\hat{f}_{AB}^{\hphantom{AB}C}\frac{\hat{J}_{C}(y)}{(z-y)},
\end{equation}
when conveniently normalized. Here, $\delta_{AB}$ is a Kronecker
delta.

In order to analyze the physical spectrum, we can start with the hatted
version of \eqref{eq:vertexJ}, defined by 
\begin{eqnarray}
V_{\widehat{J}} & = & c_{+}c_{-}P_{m}^{+}\hat{J}_{A}\hat{G}_{A}^{m}+c_{-}(\partial c_{+}-\partial c_{-})\hat{J}_{A}\hat{G}^{A}+c_{+}c_{-}\partial\hat{J}_{A}\hat{S}_{A}\nonumber \\
 &  & +c_{+}c_{-}P_{m}^{-}\hat{J}_{A}\hat{F}_{A}^{m}+c_{+}(\partial c_{+}-\partial c_{-})\hat{J}_{A}\hat{F}^{A}+c_{+}c_{-}\hat{J}_{\hat{\alpha}}\hat{\varphi}^{\hat{\alpha}}.\label{eq:vertexJhat}
\end{eqnarray}
It is easy to see that the fields appearing in this vertex operator
will satisfy essentially the same equations of motion and gauge transformations
as their counterparts in the ``$-$'' sector, albeit with one important
difference: the replacement ${\cal T}\to-{\cal T}$. By going through
the same steps as in subsection \ref{subsec:physicalspectrum}, we
find that the physical spectrum in this sector contains a ``mirror
image'' of the physical spectrum in the ``$-$'' sector, but with
opposite mass-squared.

In addition, we can build a new type of vertex operator involving
currents from both sectors. It has the form 
\begin{equation}
V_{\phi}=c_{+}c_{-}J_{a}\hat{J}_{A}\phi^{aA},\label{eq:phivertex}
\end{equation}
where $\phi^{aA}$ is a bi-adjoint scalar transforming in the adjoint
representation of both gauge groups. BRST closedness implies the equation
of motion 
\begin{equation}
\Box\phi^{aA}=0,
\end{equation}
whence $\phi^{aA}$ is a massless field.

Following the same method used in subsection \ref{subsec:EFT}, the
kinetic part of the effective action involving the group indices can
be cast as 
\begin{equation}
S^{0}=S_{J}^{0}+S_{\widehat{J}}^{0}+S_{\phi}^{0},
\end{equation}
where $S_{J}^{0}$ was given in \eqref{eq:S_J^0} and $S_{\widehat{J}}^{0}$
is its hatted analogue, and 
\begin{equation}
S_{\phi}^{0}=k\hat{k}\int{\rm d}^{d}X\{\phi_{aA}\Box\phi^{aA}\}.
\end{equation}

As for the interacting part, it clearly contains the corresponding
part in \eqref{eq:Effectiveaction-minus} and its hatted version.
Moreover, note that cubic vertices mixing the fields in $V_{J}$ with
those in $V_{\widehat{J}}$ can only appear through $\langle V_{\phi}V_{J}V_{\widehat{J}}\rangle$,
since the three-point functions involving $\langle J\hat{J}\hat{J}\rangle$
or $\langle JJ\hat{J}\rangle$ vanish. The non-vanishing three-point
functions with insertions of $V_{\phi}$ are given by: \begin{subequations}
\begin{eqnarray}
\left\langle V_{\phi}(z)V_{\phi}(y)V_{\phi}(w)\right\rangle  & = & k\hat{k}f^{abc}\hat{f}^{ABC}\left\langle \phi^{aA}\phi^{bB}\phi^{cC}\right\rangle ,\\
\left\langle V_{\phi}(z)V_{\phi}(y)V_{J}(w)\right\rangle  & = & -ik\hat{k}f_{abc}\left\langle \phi^{aA}\partial_{m}\phi^{bA}(F_{c}^{m}+G_{c}^{m})\right\rangle \nonumber \\
 &  & -k\hat{k}C_{\alpha ab}\left\langle \phi^{aA}\phi^{bA}\varphi^{\alpha}\right\rangle ,\\
\left\langle V_{\phi}(z)V_{\phi}(y)V_{\widehat{J}}(w)\right\rangle  & = & -ik\hat{k}f_{ABC}\left\langle \phi^{aA}\partial_{m}\phi^{aB}(\hat{F}_{C}^{m}+\hat{G}_{C}^{m})\right\rangle \nonumber \\
 &  & -k\hat{k}\hat{C}_{\hat{\alpha}AB}\left\langle \phi^{aA}\phi^{aB}\hat{\varphi}^{\hat{\alpha}}\right\rangle ,\\
\left\langle V_{\phi}(z)V_{J}(y)V_{\widehat{J}}(w)\right\rangle  & = & \tfrac{1}{2}k\hat{k}\eta_{mn}\left\langle (F_{a}^{m}+G_{a}^{m})(\hat{F}_{A}^{n}+\hat{G}_{A}^{n})\Box\phi^{aA}\right\rangle \nonumber \\
 &  & -k\hat{k}\eta_{mn}\left\langle \phi^{aA}\partial_{p}(F_{a}^{m}+G_{a}^{m})\partial^{p}(\hat{F}_{A}^{n}+\hat{G}_{A}^{n})\right\rangle \nonumber \\
 &  & +k\hat{k}\left\langle \phi^{aA}\partial_{n}(F_{a}^{m}+G_{a}^{m})\partial^{m}(\hat{F}_{A}^{n}+\hat{G}_{A}^{n})\right\rangle .
\end{eqnarray}
\end{subequations}

Thus, defining
\begin{equation}
\begin{array}{ccc}
\hat{A}_{A}^{m}\equiv\hat{F}_{A}^{m}+\hat{G}_{A}^{m}, &  & \hat{F}_{A}^{mn}\equiv\partial^{m}\hat{A}_{A}^{n}-\partial^{n}\hat{A}_{A}^{m}+ig\hat{f}_{ABC}\hat{A}_{B}^{m}\hat{A}_{C}^{n},\end{array}
\end{equation}
and following arguments similar to the ones given in the previous
subsection, we can write the effective action as 
\begin{equation}
S_{eff}=S[A,\varphi]+S[\hat{A},\hat{\varphi}]+S[A,\hat{A},\phi],
\end{equation}
where $S[A,\varphi]$ is the right-hand side of \eqref{eq:Effectiveaction-minus},
$S[\hat{A},\hat{\varphi}]$ is its hatted version and 
\begin{multline}
S[A,\hat{A},\phi]\equiv\int{\rm d}^{d}x\Big\{\frac{\hat{k}}{2}(D_{m}\phi^{aA})^{2}+\frac{g\hat{k}}{3!}f_{abc}\hat{f}_{ABC}\phi^{aA}\phi^{bB}\phi^{cC}+\frac{g}{2}C_{\alpha ab}\varphi^{\alpha}\phi^{aA}\phi^{bA}\\
+\frac{g}{2}\hat{C}_{\hat{\alpha}AB}\hat{\varphi}^{\hat{\alpha}}\phi^{aA}\phi^{aB}+g\phi^{aA}F_{a}^{mn}\hat{F}_{mnA}\Big\},\label{Sphi}
\end{multline}
where the covariant derivative of $\phi^{aA}$ with respect to both
gauge fields is given by 
\begin{equation}
D^{m}\phi^{aA}=\partial^{m}\phi^{aA}-igf_{abc}A_{b}^{m}\phi^{cA}-ig\hat{f}_{ABC}\hat{A}_{B}^{m}\phi^{aC}.
\end{equation}

Thus we have found the complete effective action in the gauge sector
of the model. Now we would like to make contact with the scalar extension
of the $(DF)^{2}+{\rm YM}$ theory which was introduced by Johansson
and Nohle \cite{Johansson:2017srf}. There, the group $\hat{G}$ (with
indices $A,B,\ldots$) is viewed instead as a global symmetry group.\footnote{In the context of the double-copy construction found in \cite{Azevedo:2018dgo},
this would be the heterotic string group.} In the present chiral string formulation, we can turn off the gauge
field $\hat{A}_{A}^{m}$ and the scalar $\hat{\varphi}^{\hat{\alpha}}$,
effectively taking $S[\hat{A},\hat{\varphi}]\to0$ and turning the
group $\hat{G}$ into a global symmetry at tree level. Moreover, we
are free to rescale the field $\phi$ in order to eliminate $\hat{k}$ from its kinetic term. However, a factor of $\lambda \equiv \sqrt{\hat{k}}$ would still be present in the cubic term (with $\lambda>0$).  After performing these modifications, we can
finally write the effective Lagrangian in the same form as in \cite{Johansson:2017srf}:
\begin{eqnarray}
{\cal L}_{(DF)^{2}+{\rm YM}+\phi^{3}} & = & \frac{1}{2}(D_{n}F_{a}^{mn})^{2}+\frac{1}{2}(D_{m}\varphi^{\alpha})^{2}+\frac{1}{2}(D_{m}\phi_{aA})^{2}+\frac{1}{2}m^{2}(\varphi^{\alpha})^{2}+\frac{1}{4}m^{2}(F_{a}^{mn})^{2}\nonumber \\
 &  & +\frac{g}{3}F^{3}+\frac{g}{2}C_{\alpha ab}\varphi^{\alpha}F^{mna}F_{mn}^{b}+\frac{g}{3!}d_{\alpha\beta\gamma}\varphi^{\alpha}\varphi^{\beta}\varphi^{\gamma}+\frac{g}{2}C_{\alpha ab}\varphi^{\alpha}\phi^{aA}\phi^{bA}\nonumber \\
 &  & +\frac{g\lambda}{3!}f_{abc}\hat{f}_{ABC}\phi^{aA}\phi^{bB}\phi^{cC},
\end{eqnarray}
where $m^{2}=-4{\cal T}$.

\section{Conclusion\label{sec:conclusion}}

In the first part of this work, we reexamined the bosonic chiral string, now in the sectorized interpretation, deriving a few novel results. By considering the action of the BRST operator on
the most general vertex operator, we confirmed  the physical spectrum found in \cite{Huang:2016bdd}, namely a massless level identical to that of the ordinary bosonic string and two traceless-symmetric fields $h_\pm^{mn}$ with mass-squared $m^2 = \pm4{\cal T}$. Moreover, we showed  that the
extra (massive) states  can be seen as auxiliary fields (cf. footnote \ref{nota1})
 leading to a
higher derivative gravity theory, which in the tensionless limit ($\mathcal{T}\to0$) reduces
to the recent  results of \cite{Berkovits:2018jvm} for the bosonic ambitwistor string. In \cite{Leite:2016fno}
the massive spin-$2$ states were determined to be ghosts via a $4$-point
amplitude analysis based on a ``twisted''  Kawai--Lewellen--Tye formula. This fact is manifest in    the quadratic action we construct from the vertex operator.

In the second part, we showed that the current algebra extension of
the bosonic model effectively leads to the $(DF)^{2}+\text{YM}+\phi^{3}$
Lagrangian of \cite{Johansson:2017srf}, with all its fields and couplings
coming naturally from standard string (field) theory techniques. The emergence of the higher derivative
term $(DF)^{2}$ from two vector fields of the physical spectrum 
is particularly interesting.
In addition, we would like to point out that the group constants $C_{\alpha ab}$
and $d_{\alpha\beta\gamma}$, their relations and properties emerge
naturally in our model and are valid for a generic level $k$ of the
algebra. In \cite{Johansson:2017srf}, on the other hand, such relations
are obtained by  demanding that the gluon amplitudes satisfy the Bern--Carrasco--Johansson relations \cite{Bern:2008qj}
and  our results agree  when we take $k\to0$.
This limit corresponds to a projection to the single-trace amplitude sector, which is where we expect our results to match. The multitrace sector of the worldsheet model is ``contaminated'' by the gravity theory described in section \ref{sec:bosonic-sectorized}, much like the Berkovits--Witten twistor string necessarily includes conformal gravity \cite{Witten:2003nn,Berkovits:2004jj}.

It is amusing to notice that the amplitudes of $(DF)^{2}$ theories, which enter
as double-copy constituents of (ordinary) bosonic and heterotic string amplitudes \cite{Azevedo:2018dgo},
can themselves be described in terms of a worldsheet model. Since the chiral string we studied is a closed string, it is plausible that its amplitudes can be written in  double-copy form. The latter would probably involve the Z-theory  investigated in \cite{Carrasco:2016ldy,Mafra:2016mcc,Carrasco:2016ygv}, which is related to the basis of integrals mentioned in the introduction. We plan to address this question in future work.

\

\textbf{Acknowledgments:} We would like to thank Nathan Berkovits and Fei Teng for useful discussions. TA would like to thank Marco Chiodaroli, Henrik Johansson and Oliver Schlotterer for collaboration on related topics. RLJ would like to thank the Czech Science Foundation - GA\v{C}R for financial support under the grant 19-06342Y. ML would like to thank FAPESP grant 2016/16824-0 for financial support.

\appendix

\section{Current algebra CFT\label{sec:currentalgebraCFT}}

In this appendix we will discuss some general properties of the CFT
of gauge sector of section \ref{sec:gaugesector}.

As mentioned in the text, we are using the ordering prescription \eqref{eq:ordering},
which can be understood as the product of two operators $A(z)$ and
$B(y)$ in the limit $z\to y$ with the removal of singular terms.
Note that this prescription is neither commutative nor associative:
\begin{eqnarray}
\left(A,B\right) & \neq & \left(B,A\right),\\
\left(\left(A,B\right),C\right) & \neq & \left(A,\left(B,C\right)\right).
\end{eqnarray}

The energy-momentum tensor of the algebra can be obtained using the
Sugawara construction and it is defined by 
\begin{equation}
T\equiv A\left(J_{a},J_{a}\right),\label{eq:defT}
\end{equation}
where $A$ is a numerical constant to be determined by imposing the
OPE
\begin{equation}
J_{a}(z)\,T(y)\sim\frac{J_{a}}{(z-y)^{2}}.
\end{equation}

In order to do that, we can compute first
\begin{eqnarray}
J_{a}(z)\,\left(J_{b},J_{c}\right)(y) & \sim & ik\frac{f_{abd}\delta_{dc}}{(z-y)^{3}}-f_{abd}f_{dce}\frac{J_{e}}{(w-y)^{2}}\nonumber \\
 &  & +\frac{k\delta_{ac}J_{b}}{(z-y)^{2}}+\frac{if_{acd}}{(z-y)}\left(J_{b},J_{d}\right)\nonumber \\
 &  & +\frac{k\delta_{ab}J_{c}}{(z-y)^{2}}+\frac{if_{abd}}{(z-y)}\left(J_{d},J_{c}\right).
\end{eqnarray}
It implies that
\begin{equation}
J_{a}(z)\,T(y)\sim2Ak\frac{J_{a}}{(z-y)^{2}}+Af_{acd}f_{bcd}\frac{J_{b}}{(w-y)^{2}}.
\end{equation}
Now we introduce the dual Coxeter number, $g$, defined through
\begin{equation}
f_{acd}f_{bcd}=2g\delta_{ab}.
\end{equation}
Therefore we can fix $A$ to
\begin{equation}
A=\frac{1}{2(k+g)}.
\end{equation}

Now we can compute the central charge of the model through the OPE
\begin{equation}
T(z)T(y)\sim\frac{c/2}{(z-y)^{4}}+\frac{2T}{(z-y^{2})}+\frac{\partial T}{(z-y)}.
\end{equation}
The result is
\begin{equation}
c=\frac{k\Delta}{(k+g)}.
\end{equation}
This is the central charge of the gauge sector.

\subsection*{Building additional primary operators}

One of the operators we need for the computation of $3$-point amplitudes
is related to the ordered product of two currents, $\left(J_{a},J_{b}\right)$.
Observe, however, that this product is not symmetric. In fact, we
can show that
\begin{equation}
\left(J_{a},J_{b}\right)-\left(J_{b},J_{a}\right)=if_{abc}\partial J_{c}.
\end{equation}
Therefore, we can define the operator $J_{ab}=J_{ba}$ as
\begin{eqnarray}
J_{ab} & \equiv & \tfrac{1}{2}\left(J_{a},J_{b}\right)+\tfrac{1}{2}\left(J_{b},J_{a}\right),\\
 & = & \left(J_{a},J_{b}\right)-\tfrac{i}{2}f_{ab}^{\hphantom{ab}c}\partial J_{c},
\end{eqnarray}
which can be further decomposed in two irreducible pieces: its trace,
proportional to $T$, and a traceless part.

Observe that any rank two tensor $T_{ab}$ can automatically generate
a symmetric traceless tensor $T_{(ab)}$ via a multiplication by the
projector
\[
\delta_{(ab)(cd)}\equiv\tfrac{1}{2}\delta_{ac}\delta_{bd}+\tfrac{1}{2}\delta_{ad}\delta_{bc}-\tfrac{1}{\Delta}\delta_{ab}\delta_{cd}.
\]
It acts as an identity operator for the indices $(ab)$, as
\begin{equation}
\delta_{(ab)(ef)}\delta_{(ef)(cd)}=\delta_{(ab)(cd)}.
\end{equation}
The pair $(ab)$ is an explicit realization of the index $\alpha$
introduced in section \eqref{sec:gaugesector}, labeling the field
$\varphi_{\alpha}$ of the vertex operator \eqref{eq:vertexJ}.

As it turns out, the symmetric traceless projection picks only the
primary part of the operator $\left(J_{a},J_{b}\right)$:
\begin{equation}
T(z)\,\delta_{(ab)(cd)}\left(J_{c},J_{d}\right)(y)\sim\delta_{(ab)(cd)}\frac{2\left(J_{c},J_{d}\right)}{(z-y)^{2}}+\delta_{(ab)(cd)}\frac{\partial\left(J_{c},J_{d}\right)}{(z-y)}.
\end{equation}
This is the only dimension $2$ primary operator that can be build
out of the currents $J_{a}$. In addition, we will define the operator
\begin{equation}
\Delta_{(ab)(cd)}\equiv\tfrac{1}{2}f_{ade}f_{bce}+\tfrac{1}{2}f_{ace}f_{bde}-\tfrac{2g}{\Delta}\delta_{ab}\delta_{cd},
\end{equation}
which is also symmetric and traceless in the index pairs $(ab)$ and
$(cd)$, and the power series
\begin{equation}
C_{(ab)(cd)}=\delta_{(ab)(cd)}-2\sum_{n=1}^{\infty}(-1)^{n}\frac{(2n-2)!}{(8k)^{n}n!}\left(\Delta^{n}\right)_{(ab)(cd)},
\end{equation}
satisfying
\begin{equation}
C_{(ab)(ef)}C_{(ef)(cd)}=\delta_{(ab)(cd)}+\tfrac{1}{2k}\Delta_{(ab)(cd)}.
\end{equation}
This is a realization of the Clebsch-Gordan coefficients, $C_{\alpha ab}$,
introduced earlier. By construction,
\begin{equation}
\left(C^{-1}\right)_{(ab)(cd)}\equiv\delta_{(ab)(cd)}+\sum_{n=1}^{\infty}(-1)^{n}\frac{(2n)!}{(8k)^{n}n!}\left(\Delta^{n}\right)_{(ab)(cd)}.
\end{equation}

Let us now define the dimension $2$ primary operator
\begin{equation}
J_{(ab)}\equiv\left(C^{-1}\right)_{(ab)(cd)}\left(J_{c},J_{d}\right),
\end{equation}
which satisfies the OPE
\[
J_{a}(z)\,J_{(bc)}(y)\sim2kC_{(ad)(bc)}\frac{J_{d}}{(z-y)^{2}}-\left(T_{a}\right)_{(bc)(de)}\frac{J_{(de)}}{(z-y)},
\]
where
\begin{equation}
\left(T_{a}\right)_{(bc)(de)}\equiv-2i\left(C^{-1}\right)_{(bc)(fg)}f_{afh}C_{(gh)(de)}
\end{equation}
Observe that $\left(T_{a}\right)_{(bc)(de)}$ constitutes a representation
of the group generator, as
\begin{eqnarray}
\left[T_{a},T_{b}\right]_{(de)(fg)} & \equiv & \left(T_{a}\right)_{(de)(hi)}\left(T_{b}\right)_{(hi)(fg)}-\left(T_{b}\right)_{(cd)(gh)}\left(T_{a}\right)_{(gh)(ef)}.\nonumber \\
 & = & if_{abc}\left(T_{c}\right)_{(de)(fg)}.
\end{eqnarray}
In addition, it satisfies\begin{subequations}
\begin{eqnarray}
\left(T_{a}T_{a}\right)_{(bc)(de)} & = & 2g\delta_{bd}\delta_{ce}+2g\delta_{be}\delta_{cd}-f_{bdf}f_{cef}-f_{bef}f_{cdf},\\
\left(T_{a}T_{b}\right)_{(cd)(cd)} & = & 2g(\Delta+2)\delta_{ab}.
\end{eqnarray}
\end{subequations} 

At the next conformal level, there are only two primary operators
that can be build out of $J_{a}$, defined as\begin{subequations}
\begin{eqnarray}
J_{[ab]} & \equiv & \tfrac{1}{2}\left(J_{a},\partial J_{b}\right)-\tfrac{1}{2}\left(J_{b},\partial J_{a}\right)-\tfrac{i}{3}f_{abc}\partial^{2}J_{c}+iCf_{abc}\left(J_{c},T\right),\\
J_{(abc)} & = & J_{abc}-C\left[\delta_{bc}\left(J_{a},T\right)+\delta_{ac}\left(J_{b},T\right)+\delta_{ab}\left(J_{c},T\right)\right],
\end{eqnarray}
\end{subequations}where\begin{subequations}
\begin{eqnarray}
J_{abc} & \equiv & \tfrac{1}{3}\left(J_{a},J_{bc}\right)+\tfrac{1}{3}\left(J_{b},J_{ac}\right)+\tfrac{1}{3}\left(J_{c},J_{ab}\right),\\
C & = & \frac{2(k+g)}{k\Delta+2(k+g)}.
\end{eqnarray}
\end{subequations} 

They are naturally generated in the OPE algebra. For example,
\begin{eqnarray}
J_{(bc)}(z)\,J_{(ad)}(y) & \sim & 2k^{2}\frac{\delta_{(ad)(bc)}}{(z-y)^{4}}+\tfrac{4k(k+g)}{\Delta}\delta_{(ad)(bc)}\left\{ \frac{2T}{(z-y)^{2}}+\frac{\partial T}{(z-y)}\right\} \nonumber \\
 &  & -k\left(T_{e}\right)_{(bc)(ad)}\left\{ \frac{2J_{e}}{(z-y)^{3}}+\frac{\partial J_{e}}{(z-y)^{2}}+\tfrac{1}{3}\frac{\partial^{2}J_{e}}{(z-y)}\right\} \nonumber \\
 &  & +\tfrac{1}{2}\left(C^{-1}\right)_{(ad)(gh)}\left(T_{g}T_{h}\right)_{(bc)(ef)}\left\{ \frac{2J_{(ef)}}{(z-y)^{2}}+\frac{J_{(ef)}}{(z-y)^{2}}\right\} \nonumber \\
 &  & +D_{d(bc)(ef)}\frac{J_{(aef)}}{(z-y)}+D_{a(bc)(ef)}\frac{J_{(def)}}{(z-y)}\nonumber \\
 &  & +D_{(bc)(ad)[ef]}\frac{J_{[ef]}}{(z-y)}+E_{e(bc)(ad)}\frac{\left(J_{e},T\right)}{(z-y)},
\end{eqnarray}
where $D_{a(bc)(de)}$, $D_{(ab)(cd)[ef]}$ and $E_{a(bc)(de)}$ are
given in terms of the structure constants of the group, but their
precise expression will not be needed here. The OPE above was presented
in the main text with the indices $\alpha,\beta$ in equation \eqref{eq:OPEJJ-JJ}.


\begin{thebibliography}{10}
\bibitem{Cachazo:2013hca}F.~Cachazo, S.~He and E.~Y.~Yuan, ``Scattering of Massless Particles in Arbitrary Dimensions,'' Phys.Rev.Lett.113, no. 17, 171601 (2014) doi:10.1103/PhysRevLett.113.171601 [arXiv:1307.2199 [hep-th]].

\bibitem{Cachazo:2013iea} F.~Cachazo, S.~He and E.~Y.~Yuan, ``Scattering of Massless Particles: Scalars, Gluons and Gravitons,'' JHEP 1407, 033 (2014) doi:10.1007/JHEP07(2014)033 [arXiv:1309.0885 [hep-th]].

\bibitem{Mason:2013sva}L.~Mason and D.~Skinner, ``Ambitwistor strings and the scattering equations,'' JHEP 1407, 048 (2014) doi:10.1007/JHEP07(2014)048 [arXiv:1311.2564 [hep-th]].

\bibitem{Berkovits:2013xba}N.~Berkovits, ``Infinite Tension Limit of the Pure Spinor Superstring,''   JHEP {\bf 1403}, 017 (2014)   doi:10.1007/JHEP03(2014)017   [arXiv:1311.4156 [hep-th]].

\bibitem{Cachazo:2014nsa}F.~Cachazo, S.~He and E.~Y.~Yuan, ``Einstein-Yang-Mills Scattering Amplitudes From Scattering Equations,''   JHEP {\bf 1501}, 121 (2015)   doi:10.1007/JHEP01(2015)121   [arXiv:1409.8256 [hep-th]].

\bibitem{Cachazo:2014xea}F.~Cachazo, S.~He and E.~Y.~Yuan, ``Scattering Equations and Matrices: From Einstein To Yang-Mills, DBI and NLSM,''   JHEP {\bf 1507}, 149 (2015)   doi:10.1007/JHEP07(2015)149   [arXiv:1412.3479 [hep-th]].

\bibitem{Casali:2015vta} E.~Casali, Y.~Geyer, L.~Mason, R.~Monteiro and K.~A.~Roehrig, ``New Ambitwistor String Theories,'' JHEP 1511, 038 (2015) doi:10.1007/JHEP11(2015)038 [arXiv:1506.08771 [hep-th]].

\bibitem{Bandos:2014lja}I.~Bandos, ``Twistor/ambitwistor strings and null-superstrings in spacetime of D=4, 10 and 11 dimensions,''   JHEP {\bf 1409}, 086 (2014)   doi:10.1007/JHEP09(2014)086   [arXiv:1404.1299 [hep-th]].

\bibitem{Siegel:2015axg}W.~Siegel, ``Amplitudes for left-handed strings,''   arXiv:1512.02569 [hep-th].

\bibitem{Casali:2016atr}E.~Casali and P.~Tourkine, ``On the null origin of the ambitwistor string,''   JHEP {\bf 1611}, 036 (2016)   doi:10.1007/JHEP11(2016)036   [arXiv:1606.05636 [hep-th]].

\bibitem{Huang:2016bdd}Y.~t.~Huang, W.~Siegel and E.~Y.~Yuan, ``Factorization of Chiral String Amplitudes,''   JHEP {\bf 1609}, 101 (2016)   doi:10.1007/JHEP09(2016)101   [arXiv:1603.02588 [hep-th]].

\bibitem{Jusinskas:2016qjd}R.~L.~Jusinskas, ``Notes on the ambitwistor pure spinor string,''   JHEP {\bf 1605}, 116 (2016)   doi:10.1007/JHEP05(2016)116   [arXiv:1604.02915 [hep-th]].

\bibitem{Chandia:2015sfa}O.~Chandia and B.~C.~Vallilo, ``Ambitwistor pure spinor string in a type II supergravity background,''   JHEP {\bf 1506}, 206 (2015)   doi:10.1007/JHEP06(2015)206   [arXiv:1505.05122 [hep-th]].

\bibitem{Azevedo:2017yjy}T.~Azevedo and R.~L.~Jusinskas, ``Connecting the ambitwistor and the sectorized heterotic strings,''   JHEP {\bf 1710}, 216 (2017)   doi:10.1007/JHEP10(2017)216   [arXiv:1707.08840 [hep-th]].

\bibitem{Berkovits:2018jvm}N.~Berkovits and M.~Lize, ``Field theory actions for ambitwistor string and superstring,''   JHEP {\bf 1809}, 097 (2018)   doi:10.1007/JHEP09(2018)097   [arXiv:1807.07661 [hep-th]].

\bibitem{Zwiebach:1992ie}B.~Zwiebach,   ``Closed string field theory: Quantum action and the B-V master equation,''   Nucl.\ Phys.\ B {\bf 390}, 33 (1993)   doi:10.1016/0550-3213(93)90388-6   [hep-th/9206084].

\bibitem{Johansson:2017srf}H.~Johansson and J.~Nohle,   ``Conformal Gravity from Gauge Theory,''   arXiv:1707.02965 [hep-th].

\bibitem{Bern:2008qj}Z.~Bern, J.~J.~M.~Carrasco and H.~Johansson,   ``New Relations for Gauge-Theory Amplitudes,''   Phys.\ Rev.\ D {\bf 78}, 085011 (2008)   doi:10.1103/PhysRevD.78.085011   [arXiv:0805.3993 [hep-ph]].

\bibitem{Azevedo:2017lkz} T.~Azevedo and O.~T.~Engelund,   ``Ambitwistor formulations of R$^{2}$ gravity and (DF)$^{2}$ gauge theories,''   JHEP {\bf 1711}, 052 (2017)   doi:10.1007/JHEP11(2017)052   [arXiv:1707.02192 [hep-th]].

\bibitem{Azevedo:2018dgo}T.~Azevedo, M.~Chiodaroli, H.~Johansson and O.~Schlotterer,  ``Heterotic and bosonic string amplitudes via field theory,''   JHEP {\bf 1810}, 012 (2018)   doi:10.1007/JHEP10(2018)012   [arXiv:1803.05452 [hep-th]].

\bibitem{He:2018pol} S.~He, F.~Teng and Y.~Zhang, ``String amplitudes from field-theory amplitudes and vice versa,''   Phys.\ Rev.\ Lett.\  {\bf 122}, no. 21, 211603 (2019)   doi:10.1103/PhysRevLett.122.211603   [arXiv:1812.03369 [hep-th]].

\bibitem{He:2019drm} S.~He, F.~Teng and Y.~Zhang, ``String Correlators: Recursive Expansion, Integration-by-Parts and Scattering Equations,''   arXiv:1907.06041 [hep-th].

\bibitem{Mafra:2011nv}
C.~R. Mafra, O.~Schlotterer and S.~Stieberger, \emph{{Complete N-Point
  Superstring Disk Amplitude I. Pure Spinor Computation}},
  {\emph{Nucl. Phys.}
  {\bf B873} (2013) 419--460}, [arXiv:1106.2645 [hep-th]].

\bibitem{Hohm:2013jaa}O.~Hohm, W.~Siegel and B.~Zwiebach, ``Doubled $\alpha'$-geometry,''   JHEP {\bf 1402}, 065 (2014)   doi:10.1007/JHEP02(2014)065   [arXiv:1306.2970 [hep-th]].

\bibitem{Lee:2017utr} 
  K.~Lee, S.~J.~Rey and J.~A.~Rosabal,
  JHEP {\bf 1711}, 172 (2017)
  doi:10.1007/JHEP11(2017)172
  [arXiv:1708.05707 [hep-th]].

\bibitem{Casali:2017zkz} 
  E.~Casali, Y.~Herfray and P.~Tourkine,
  ``The complex null string, Galilean conformal algebra and scattering equations,''
  JHEP {\bf 1710}, 164 (2017)
  doi:10.1007/JHEP10(2017)164
  [arXiv:1707.09900 [hep-th]].


\bibitem{Leite:2016fno} M.~M.~Leite and W.~Siegel, ``Chiral Closed strings: Four massless states scattering amplitude,'' JHEP {\bf 1701}, 057 (2017)   doi:10.1007/JHEP01(2017)057   [arXiv:1610.02052 [hep-th]].

\bibitem{Witten:2003nn} 
  E.~Witten,   ``Perturbative gauge theory as a string theory in twistor space,''   Commun.\ Math.\ Phys.\  {\bf 252}, 189 (2004)
  doi:10.1007/s00220-004-1187-3   [hep-th/0312171].

\bibitem{Berkovits:2004jj}   N.~Berkovits and E.~Witten, ``Conformal supergravity in twistor-string theory,''   JHEP {\bf 0408} (2004) 009   doi:10.1088/1126-6708/2004/08/009   [hep-th/0406051].
 
\bibitem{Carrasco:2016ldy} J.~J.~M.~Carrasco, C.~R.~Mafra and O.~Schlotterer, ``Abelian Z-theory: NLSM amplitudes and $\alpha$'-corrections from the open string,''   JHEP {\bf 1706}, 093 (2017)   doi:10.1007/JHEP06(2017)093   [arXiv:1608.02569 [hep-th]].

\bibitem{Mafra:2016mcc} C.~R.~Mafra and O.~Schlotterer, ``Non-abelian $Z$-theory: Berends-Giele recursion for the $\alpha'$-expansion of disk integrals,''   JHEP {\bf 1701}, 031 (2017)   doi:10.1007/JHEP01(2017)031   [arXiv:1609.07078 [hep-th]].

\bibitem{Carrasco:2016ygv} J.~J.~M.~Carrasco, C.~R.~Mafra and O.~Schlotterer, ``Semi-abelian Z-theory: NLSM$+\phi^{3}$ from the open string,''   JHEP {\bf 1708}, 135 (2017)   doi:10.1007/JHEP08(2017)135   [arXiv:1612.06446 [hep-th]].

\


\end{thebibliography}
\end{document}